\documentclass[letterpaper,12pt,leqno]{article}
\usepackage{paper,math}
\pdfoutput=1
\renewcommand{\itshape}{\fontfamily{LinuxLibertineT-LF}\fontshape{it}\selectfont}
\renewcommand{\scshape}{\fontfamily{LinuxLibertineT-LF}\fontshape{sc}\selectfont}
\def\bib{paper.bib}
\def\pdf{figures.pdf}
\published{Journal of the European Economic Association}{https://www.pascalmichaillat.org/8.html}
\hypersetup{pdftitle={Pricing under Fairness Concerns}}
\hypersetup{pdfauthor={Erik Eyster, Kristof Madarasz, Pascal Michaillat}}

\begin{document}

\title{Pricing under Fairness Concerns}
\author{Erik Eyster, Krist\'{o}f Madar\'{a}sz, Pascal Michaillat
\thanks{Eyster: University of California--Santa Barbara. Madar\'{a}sz: London School of Economics. Michaillat: Brown University. We thank George Akerlof, Roland Benabou, Daniel Benjamin, Joaquin Blaum, Arpita Chatterjee, Varanya Chaubey, Olivier Coibion, Stephane Dupraz, Gauti Eggertsson, John Friedman, Xavier Gabaix, Nicola Gennaioli, Yuriy Gorodnichenko, David Laibson, John Leahy, Bobak Pakzad-Hurson, Matthew Rabin, Ricardo Reis, David Romer, Emmanuel Saez, Klaus Schmidt, Jesse Shapiro, Andrei Shleifer, Joel Sobel, and Silvana Tenreyro for helpful discussions and comments. This work was supported by the Institute for Advanced Study.}}
\date{August 2020}

\begin{titlepage}\maketitle\begin{abstract}

This paper proposes a theory of pricing premised upon the assumptions that customers dislike unfair prices---those marked up steeply over cost---and that firms take these concerns into account when setting prices. Since they do not observe firms' costs, customers must extract costs from prices. The theory assumes that customers infer less than rationally: when a price rises due to a cost increase, customers partially misattribute the higher price to a higher markup---which they find unfair. Firms anticipate this response and trim their price increases, which drives the passthrough of costs into prices below one: prices are somewhat rigid. Embedded in a New Keynesian model as a replacement for the usual pricing frictions, our theory produces monetary nonneutrality: when monetary policy loosens and inflation rises, customers misperceive markups as higher and feel unfairly treated; firms mitigate this perceived unfairness by reducing their markups; in general equilibrium, employment rises. The theory also features a hybrid short-run Phillips curve, realistic impulse responses of output and employment to monetary and technology shocks, and an upward-sloping long-run Phillips curve.

\end{abstract}\end{titlepage}\section{Introduction}

Prices are neither exactly fixed nor fully responsive to cost shocks \cp{CS12,DGK16,CCW17,GSW19}. Such price rigidity has first-order importance by determining how economic shocks percolate through the economy as well as the effectiveness of different policy responses.

Asked why they show such restraint when setting prices, firm managers explain that they endeavor to avoid alienating customers, who balk at paying prices that they regard as unfair \cp{BCL98}. Yet theories of price rigidity never include fairness considerations \cp{B90,MR10}, with the notable exception of the theory by \ct{R05}, which calls attention to the role of fairness in pricing.\footnote{Fairness has received more attention in other contexts: \ct{A82}, \ct{AY90}, and \ct{B15} add fairness to labor-market models; \ct{R93}, \ct{FS99}, and \ct{CR02} to game-theoretic models; and \ct{FKS07} to contract-theoretic models. For surveys of the fairness literature, see \ct{FG00}, \ct{S05}, and \ct{FGZ09}.} Due to its innovative nature, however, that theory is somewhat difficult to analyze or use in other frameworks (see Section~\ref{s:literature}).

This paper develops a theory of pricing that incorporates the fairness concerns exhibited by firms and their customers and uses such concerns to generate the price rigidity observed in the data. The theory permits closed-form expressions for price markups and passthroughs, as well as a set of comparative statics. It also transfers easily to other frameworks: here, we port it to a New Keynesian model to study the macroeconomic implications.

The first element of our theory is that customers dislike paying prices marked up heavily over marginal costs because they find these prices unfair, and that firms understand this. This assumption draws upon evidence from numerous surveys of consumers and firms, our own survey of French bakers, and religious and legal texts (Section~\ref{s:evidence}). We formalize this assumption by weighting each unit of consumption in the utility function by a fairness factor which depends upon the markup that customers perceive firms as charging: the fairness factor decreases in the perceived markup---higher markups seem less fair---and is concave---people respond more strongly to increases in markups than to decreases. 

Customers who cannot observe firms' costs estimate these costs from prices, and then use their estimates to evaluate firms' fairness. The second element of our theory is that customers update their beliefs about marginal costs less than rationally. First, customers underinfer marginal cost from price: their beliefs depend upon some anchor, which may be their prior expectation of marginal cost. Second, insofar as customers do update their beliefs about marginal cost from price, they engage in a form of proportional thinking by estimating a marginal cost that is proportional to price. We dub this pair of assumptions subproportional inference. They draw upon evidence that during inflationary periods people seem to underinfer increases in nominal costs, and more generally that people tend to infer less than they should about others' private information from others' actions (see also Section~\ref{s:evidence}).

We begin our analysis by embedding these psychological elements into a model of monopoly pricing (Section~\ref{s:monopoly}). The monopolist's price features a markup that decreases in the price elasticity of demand. We assume a standard utility function, with the property that customers without fairness concerns would exhibit constant price elasticity of demand. 

Absent fairness concerns, the monopolist would set a constant markup that produces flexible prices, which move proportionally to marginal costs. If customers cared about fairness and rationally inverted the price to uncover the hidden marginal cost, the same pricing rule would be an equilibrium. Indeed, when price increases by $x \%$, customers correctly infer that marginal cost has increased by $x \%$, and therefore that the markup has not changed. Since the price change does not change the perceived markup, the price elasticity of demand does not change, and neither does the markup. 

Once fairness concerns and subproportional inference come together, however, pricing changes. First, demand is more price elastic than it would be otherwise, leading the monopoly to lower its markup. Indeed, demand decreases in price not only through the standard channel, but also through a fairness channel. Customers who see a higher price attribute it partially to a higher marginal cost and partially to a higher markup---which they find unfair. Thus the higher price lowers their marginal utility of consumption, which further decreases demand.

Second, demand is more price elastic at higher perceived markups, creating price rigidity. Following a price increase spurred by a cost increase, customers underappreciate the increase in marginal cost and partially misattribute the higher price to higher markup. Since the fairness factor is decreasing and concave in the perceived markup, it is more elastic at higher perceived markups; this property translates to demand. Facing a more elastic demand after the cost increase, the monopoly reduces its markup. As a result, the price increases less than proportionally to the underlying marginal cost. This mild form of price rigidity agrees with the passthroughs of marginal-cost shocks into prices estimated in empirical studies.

Price rigidity plays a central role in many macroeconomic models. To illustrate how our theory can be embedded into such models, and to develop its implications, we substitute it for the usual pricing frictions in a New Keynesian model (Section~\ref{s:nk}). Again we assume that customers infer less than they should about marginal cost from price. In the dynamic model, subproportional inference means that in each period, customers average their previous-period beliefs about marginal costs with beliefs that are proportional to current prices.

The macroeconomic model makes several realistic predictions. First, monetary policy is nonneutral in the short run: it affects output and employment. This property arises through the same channel as in the monopoly model: expansionary monetary policy raises prices and nominal marginal costs; customers partially misattribute higher prices to higher markups, which they perceive as unfair; as a result, the price elasticities of the demands for goods rise; firms respond by reducing markups, thus stimulating the economy. Second, the New Keynesian Phillips curve is hybrid: it links current employment not only to current and to expected future inflation, but also to past inflation. The reason is that consumers form backward-looking beliefs about marginal costs, which forces firms to account for both past and future inflation when setting prices. Third, the model yields reasonable impulse responses to monetary shocks and to technology shocks when the parameters governing fairness concerns and subproportional inference are calibrated to match the microevidence on cost passthroughs. In particular, the impulse responses of employment are hump-shaped. Fourth, monetary policy is nonneutral in the long run: higher steady-state inflation leads to higher steady-state employment; that is, the long-run Phillips curve is nonvertical.

Our macroeconomic model is also consistent with survey evidence that inflation angers people---who attribute it to commercial greed---whereas deflation pleases people. In our model,  because people partially misattribute higher prices to higher markups, inflation leads them to perceive transactions as less fair, generating disutility. Conversely, deflation leads people to misperceive markups as lower and deem transactions more fair, generating utility.

\section{Related Literature}\label{s:literature}

\ct{R05} developed the first theory of price rigidity based on fairness considerations.\footnote{\ct{R11} explores other implications of fairness for pricing, such as price discrimination.} Customers in his model care about firms' altruism, which they evaluate following every price change. They purchase a normal amount from a firm unless they can reject the hypothesis that the firm is sufficiently altruistic, in which case they withhold demand entirely to lower the firm's profits. Profit-maximizing firms react to the discontinuity in demand by refraining from passing on small cost increases, creating price stickiness. Consumers err in equilibrium by mistaking purely selfish firms as altruistic. 

We depart from \name{R05}'s discontinuous, buy-normally-or-buy-nothing formulation to one in which customers continuously reduce demand as the unfairness of the transaction increases. Our continuous formulation seems more realistic and offers greater tractability. The tractability allows for closed-form expressions for the markup and passthrough, and thus to obtain a range of comparative statics. The tractability also allows us to embed our pricing theory into a New Keynesian model, to calibrate the theory's parameters based on microevidence, and to perform standard simulations.

More broadly, our approach to fairness differs from the popular social-preferences approach, both the intention-based model of \ct{R93} and the distribution-based model of \ct{FS99}. Like \name{R05}'s model, these models predict that consumers endeavor to harm firms by withholding demand to lower profits in certain circumstances---namely when the firm treats consumers unkindly \cp{R93}, and when the firm receives a higher payoff than consumers \cp{FS99}. In our model, by contrast, because customers simply do not savor unfairly priced goods, they withhold demand irrespective of any harm to the firm. An advantage of our approach, which appears in its macroeconomic application, is that fairness continues to matter in general equilibrium. This is not the case with many social preferences: when people's utility can be written as a separable function of their own and others' allocations, social preferences do not affect general-equilibrium prices or allocations \cp{DHK11,S07}.

\section{Microevidence Supporting the Assumptions}\label{s:evidence}

This section presents microevidence in support of the assumptions underlying our theory. First, we show that people care about the fairness of prices, and assess prices that include low markups over marginal costs as fair. Second, we document that people erroneously infer marginal costs from prices and thus misperceive markups. Finally, we show that firms account for customers' fairness concerns when setting prices.

\subsection{Customers' Fairness Concerns}

Our theory assumes that customers find a price unfair when it entails a high markup over marginal cost, and that they dislike such prices. Here we review evidence supporting this assumption. 

\paragraph{Price Increases Due to Higher Demand} Our assumption implies that people find price increases unjustified by cost increases to be unfair. In a survey of Canadian residents, \ct[p.~729]{KKT86} document this pattern. They describe the following situation: ``A hardware store has been selling snow shovels for \$15. The morning after a large snowstorm, the store raises the price to \$20.'' Among 107 respondents, only 18\% regard this pricing as acceptable, whereas 82\% regard it as unfair.

Subsequent studies confirm and refine \name{KKT86}'s results. For example, in a survey of 1,750 households in Switzerland and Germany, \ct{FP93} confirm that customers dislike price increases that involve increased markups; so too do \ct{SBK91} in a comparative survey of 391 respondents in Russia and 361 in the United States.

The snow-shovel evidence leaves open the possibility that people find the price increase unfair simply because it occurs during a period of hardship. To address this question, \ct{M95} asks 72 students at a Florida university about price increases following an ordinary increase in demand versus those following a hardship-driven increase in demand. While more find price increases in the hardship environment unfair (86\% versus 69\%), a substantial majority in each case perceive the price increase as unfair.

\paragraph{Price Increases Due to Higher Costs}  Conversely, our fairness assumption suggests that customers tolerate price increases following cost increases so long as the markup remains constant. \ct[pp.~732--733]{KKT86} identify this pattern: ``Suppose that, due to a transportation mixup, there is a local shortage of lettuce and the wholesale price has increased. A local grocer has bought the usual quantity of lettuce at a price that is 30 cents per head higher than normal. The grocer raises the price of lettuce to customers by 30 cents per head.'' Among 101 respondents, 79\% regard the pricing as acceptable, and only 21\% find it unfair. In a survey of 307 Dutch individuals, \ct[Table~2]{GDG08} also find that price increases following cost increases are fair, while those following demand increases are not. 

\paragraph{Price Decreases Allowed by Lower Costs}  Our assumption equally implies that it is unfair for firms not to pass along cost decreases. \ct[p.~734]{KKT86} find milder support for this reaction. They describe the following situation: ``A small factory produces tables and sells all that it can make at \$200 each. Because of changes in the price of materials, the cost of making each table has recently decreased by \$20. The factory does not change its price of tables.'' Only 47\% of respondents find this unfair, despite the elevated markup. 

Subsequent studies, however, find that people do expect price reductions after cost reductions. \ct{KDU91} survey 189 US business students, asking them to consider the following scenario: ``A department store has been buying an oriental floor rug for \$100. The standard pricing practice used by department stores is to price floor rugs at double their cost so the selling price of the rug is \$200. This covers all the selling costs, overheads and includes profit. The department store can sell all of the rugs that it can buy. Suppose because of exchange rate changes the cost of the rug rises from \$100 to \$120 and the selling price is increased to \$220. As a result of another change in currency exchange rates, the cost of the rug falls by \$20 back to \$100.'' Then two alternative scenarios were evaluated: ``The department store continues to sell the rug for \$220'' compared to ``The department store reduces the price of  the rug to \$200.'' The latter was judged significantly fairer: the fairness rating was $+2.3$ instead of $-0.4$ (where $-3$ is extremely unfair and $+3$ extremely fair). Similarly, in survey of US respondents, \ct[Table~6]{K01} finds that if a factory that sells a table at \$150 locates a supplier charging \$20 less for materials, the new fair price is \$138, well below \$150.

\paragraph{Norms about Markups}  Religious and legal texts written over the ages display a long history of norms regarding markups---which suggests that people deeply care about markups. For example, Talmudic law specifies that the highest fair and allowable markup when trading essential items is 20\% of the production cost, or one-sixth of the final price \cp[p.~198]{F84}. 

Another example comes from 18th-century France, where local authorities fixed bread prices by publishing ``fair'' prices in official decrees. In the city of Rouen, for instance, the official bread prices took the costs of grain, rent, milling, wood, and labor into account, and granted a ``modest profit'' to the baker \cp[p.~36]{M99}. Thus, officials fixed the markup that bakers could charge. Even today, French bakers attach such importance to convincing their customers of fair markups that their trade union decomposes the cost of bread and the rationale for any price rise into minute detail (\url{https://perma.cc/GQ28-JMFC}). 

Two more examples come from regulation in the United States. First, return-on-cost regulation for public utilities, which limits the markups charged by utilities, has been justified not only on efficiency grounds but also on fairness grounds \cp{Z85,JM01}. Second, most US states have anti-price-gouging legislation that limits the prices that firms can charge in periods of upheaval (such as an epidemic). But by exempting price increases justified by higher costs, the legislation only outlaws price increases caused by higher markups \cp[pp.~74--77]{R09}.

\paragraph{Fairness and Willingness to Pay}  We assume that customers who purchase a good at an unfair price derive less utility from consuming it; as a result, unfair pricing reduces willingness to pay. Substantial evidence documents that unfair prices make customers angry, and more generally that unfair outcomes trigger feelings of anger \cp[pp.~60--64]{R09}. A small body of evidence also suggests that customers reduce purchases when they feel unfairly treated. In a telephone survey of 40 US consumers, \ct{UMD89} explore---by looking at a 25-cent ATM surcharge---whether a price increase justified by a cost increase is perceived as more fair than an unjustified one, and whether fairness perceptions affect customers' propensity to buy. While 58\% of respondents judge the introduction of the surcharge fair when justified by a cost increase, only 29\% judge it fair when not justified (Table~1, panel~B). Moreover, those people who find the surcharge unfair are indeed more likely to switch banks  (52\% versus 35\%; see Table~1, panel~C). Similarly, \ct{PF95} present survey and laboratory evidence that customers who find a firm's actions unfair tend to reduce their purchases with that firm.

\paragraph{Fixed Costs}  A natural question is whether the fair price would differ from a fair markup over marginal cost for businesses that have significant fixed costs. The literature almost exclusively reports on experiments with marginal costs and without fixed costs, so we cannot say how people would incorporate fixed costs into the fair price. Anecdotal evidence, however, points towards people caring directly about marginal costs. Ride-sharers outraged by Uber's surge pricing seldom mention that Uber has never reported a profit. Likewise, consumers deplore pharmaceutical companies for selling pills with very low marginal costs at high markups, without mention of R\&D expenses. In light of the evidence, we focus on marginal costs and abstract from fixed costs in the analysis.

\subsection{Subproportional Inference of Costs}

Because customers do not observe firms' marginal costs, their fairness perceptions depend upon their estimates of these costs. Since customers cannot easily learn about hidden costs, however, they are prone to develop mistaken beliefs. To describe such misperceptions, we assume subproportional inference. First, consumers underinfer marginal costs from prices: their beliefs depend too much upon some anchor. Second, insofar as consumers do update their beliefs about costs from prices, they engage in a form of proportional thinking by estimating marginal costs that are proportional to prices. We now review evidence that supports this pair of assumptions.

\paragraph{Underinference in General}  Numerous experimental studies establish that people underinfer other people's information from those other people's actions \cp{E19}. In the context of bilateral bargaining with asymmetric information, \ct{SB85}, \ct{HS94}, \ct{CP11}, and others show that bargainers underappreciate the adverse selection in trade. The papers collected in \ct{KL02} present evidence that bidders underattend to the winner's curse in common-value auctions. In a metastudy of social-learning experiments, \ct{W10} finds that subjects behave as if they underinfer their predecessors' private information from their actions. In a voting experiment, \ct{EV14} show that people underinfer others' private information from their votes. Subproportional inference encompasses such underinference.

\paragraph{Underinference from Prices}  \ct{SDT97} report survey evidence that points at underinference in the context of pricing. They presented 362 people in New Jersey with the following thought experiment: ``Changes in the economy often have an effect on people's financial decisions. Imagine that the US experienced unusually high inflation which affected all sectors of the economy. Imagine that within a six-month period all benefits and salaries, as well as the prices of all goods and services, went up by approximately 25\%. You now earn and spend 25\% more than before. Six months ago, you were planning to buy a leather armchair whose price during the 6-month period went up from \$400 to \$500. Would you be more or less likely to buy the armchair now?'' The higher prices were distinctly aversive: while 55\% of respondents were as likely to buy as before and 7\% were more likely, 38\% of respondents were less likely to buy (p.~355). Our model makes this prediction. While consumers who update subproportionally recognize that higher prices signal higher marginal costs, they stop short of rational inference. Consequently, consumers perceive markups to be higher when prices are higher. These consumers deem today's transaction less fair, so they have a lower willingness to pay for the armchair.

\paragraph{Proportional Thinking}  A small body of evidence documents that people think proportionally, even in settings that do not call for proportional thinking \cp{BRS15}. In particular, \ct{T80} and \ct{TK81} demonstrate that people's willingness to invest time in lowering the price of a good by a fixed dollar amount depends negatively upon the good's price. Rather than care about the absolute savings, people appear to care about the proportional savings. Someone who thinks about a price discount not in absolute terms but as a proportion of the purchase price may think about marginal cost not in absolute terms but rather as a percentage of price. If so, then the simplest assumption is that, insofar as the person infers marginal cost from price, she infers a marginal cost proportional to price.

\subsection{Firms' Fairness Concerns}

\begin{table}[t]
\caption{Description of firm surveys about pricing}
\footnotesize\begin{tabular*}{\textwidth}{@{\extracolsep{\fill}}lccc}\toprule
Survey & Country & Period & Number of firms\\ 
\midrule 
\ct{BCL98} & United States (US) & 1990--1992 & 200\\ 
\ct{HWY00} & United Kingdom (GB) & 1995 & 654 \\ 
\ct{AFH05} & Sweden (SE) & 2000 & 626 \\ 
\ct{NHT00} & Japan (JP) & 2000  & 630 \\
\ct{AKW06} & Canada (CA) & 2002--2003 & 170 \\ 
\ct{KBS05} & Austria (AT) & 2004 & 873 \\ 
\ct{AD05} & Belgium (BE) & 2004 & 1,979 \\ 
\ct{LR04} & France (FR) & 2004 & 1,662 \\ 
\ct{LM06} & Luxembourg (LU) & 2004 & 367 \\ 
\ct{HS06} & Netherlands (NL) & 2004 & 1,246 \\ 
\ct{M06} & Portugal (PT) & 2004 & 1,370\\ 
\ct{AH05} & Spain (ES) & 2004 & 2,008\\ 
\ct{LNW08} & Norway (NO)  & 2007 & 725 \\
\ct{OPV11} & Iceland (IS) &  2008 & 262 \\
\bottomrule\end{tabular*}
\label{t:survey}\end{table}

In our model, firms pay great attention to fairness when setting prices. This seems to hold true in the real world: firms identify fairness as a major concern in price setting.

\paragraph{Surveys of Firms}  Following \ct{BCL98}, researchers have surveyed more than 12,000 firms across developed economies about their pricing practices (Table~\ref{t:survey}). The typical study asks managers to evaluate the relevance of different pricing theories from the economics literature to explain their own pricing, in particular price rigidity. Amongst the theories that the managers deem most important, some version of fairness invariably appears, often called ``implicit contracts'' and described as follows: ``firms tacitly agree to stabilize prices, perhaps out of fairness to customers.'' Indeed, fairness appeals to firms more than any other theory, with a median rank of 1 and a mean rank of $1.9$ (Table~\ref{t:ranking}). The second most popular explanation for price rigidity takes the form of nominal contracts---prices do not change because they are fixed by contracts: it has a median rank of 3 and a mean rank of $2.6$. Two common macroeconomic theories of price rigidity---menu costs and information delays---do not resonate at all with firms, who rank them amongst the least popular theories, with mean and median ranks above~9.

\begin{sidewaystable}[p]
\caption{Ranking of pricing theories in firm surveys}
\footnotesize\begin{tabular*}{\textwidth}{@{\extracolsep{\fill}}lcccccccccccccccc}\toprule 
&\multicolumn{14}{c}{Country of survey} & \multicolumn{2}{c}{Overall rank} \\ 
\cmidrule{2-15}\cmidrule{16-17}  
Theory 		&US &GB &SE &JP &CA &AT &BE &FR &LU &NL &PT &ES &NO &IS & Median & Mean  \\
\midrule 
Implicit contracts&4 & 5 & 1 & 1 & 2 & 1 & 1 & 4 & 1 & 2 & 1 & 1 & 2 & 1 & 1  & $1.9$\\ 
Nominal contracts &5 &1 & 3 & 3 & 3 & 2 & 2 & 3 & 3 & 1 & 5 & 3 & 1 & 2 & 3 & $2.6$\\ 
Coordination failure&1 & 3& 4 & 1& 5 & 5 & 5 & 2 & 9 & 4 & 2 & 2 & 3 & 4 & $3.5$&$3.6$ \\
Pricing points & 8 & 4 & 7 & 4 &-- &10 &13 & 8 &10 & 7 &11 & 6 & 4 & 5 & 7  & $7.5$  \\
Menu costs & 6 &11 &11 & 7 &10 & 7 &15 &10 &13 & 8 &10 & 7 & 6 & 6 & 9  & $9.1$  \\
Information delays & 11&-- &13 &-- &11 & 6 &14 &--&15 &--& 8 & 9 & 5 &--& 11& $10.2$ \\ 
\bottomrule\end{tabular*}
\note{Survey respondents rated the relevance of several pricing theories to explain price rigidity at their own firm. The table ranks common theories amongst the alternatives. \ct[Table~5.1]{BCL98} describes the theories as follows (with wording that varies slightly across surveys): ``implicit contracts'' stands for ``firms tacitly agree to stabilize prices, perhaps out of fairness to customers''; ``nominal contracts'' stands for ``prices are fixed by contracts''; ``coordination failure'' stands for two closely related theories, which are investigated in separate surveys: ``firms hold back on price changes, waiting for other firms to go first'' and ``the price is sticky because the company loses many customers when it is raised, but gains only a few new ones when the price is reduced'' (which is labeled ``kinked demand curve''); ``pricing points'' stands for ``certain prices (like \$9.99) have special psychological significance''; ``menu costs'' stands for  ``firms incur costs of changing prices''; ``information delays'' stands for two closely related theories, which are investigated in separate surveys: ``hierarchical delays slow down decisions'' and ``the information used to review prices is available infrequently.'' The rankings of the theories are reported in Table~5.2 in \ct{BCL98}; Table~3 in \ct{HWY00}; Table~4 in \ct{AFH05}; Chart~14 in \ct{NHT00}; Table~8 in \ct{AKW06}; Table~5 in \ct{KBS05}; Table~18 in \ct{AD05}; Table~6.1 in \ct{LR04}; Table~8 in \ct{LM06}; Table~10 in \ct{HS06}; Table~4 in \ct{M06}; Table~5 in \ct{AH05}; Chart~26 in \ct{LNW08}; and Table~17 in \ct{OPV11}.}
\label{t:ranking}\end{sidewaystable}

Firms also understand that customers bristle at unfair markups. According to \ct[pp.~153--157]{BCL98}, 64\% of firms say that customers do not tolerate price increases after demand increases, while 71\% of firms say that customers do tolerate price increase after cost increases. And firms describe the norm for fair pricing as a constant markup over marginal cost. Based on a survey of businessmen in the United Kingdom, \ct[p.~19]{HH39} report that the ``the `right' price, the one which `ought' to be charged'' is widely perceived to be a markup (generally, 10\%) over average cost. \ct[p.~362]{O75} also observes in discussions with business people that ``empirically, the typical standard of fairness involves cost-oriented pricing with a markup.''

\paragraph{Survey of French Bakers} To better understand how firms incorporate fairness into their pricing decisions, we interviewed 31 bakers in France in 2007. The French bread market makes a good case study because the market is large; bakers set their prices freely; and French people care enormously about bread.\footnote{In 2005, bakeries employed 148,000 workers, for a yearly turnover of $3.2$ billion euros (\url{https://perma.cc/V679-UFE8}). Since 1978, French bakers have been free to set their own prices, except during the inflationary period 1979--1987 when price ceilings and growth caps were imposed. For centuries, bread prices caused major social upheaval in France. \ct[p.~35]{M99} explains that before the French Revolution, ``affordable bread prices underlay any hopes for urban tranquility.'' During the Flour War of 1775, mobs chanted ``if the price of bread does not go down, we will exterminate the king and the blood of the Bourbons''; following these riots, ``under intense pressure from irate and nervous demonstrators, the young governor of Versailles had ceded and fixed the price `in the King's name' at two sous per pound, the mythohistoric just price inscribed in the memory of the century'' \cp[p.~12]{K96}.} We sampled bakeries in Aix-en-Provence, Grenoble, Paimpol, and Paris. The interviews reveal that bakers are guided by norms of fairness when they adjust prices to preserve customer loyalty. In particular, cost-based pricing is widely used. Bakers only raise the price of bread in response to increases in the cost of flour, utilities, or wages. They refuse to increase prices in response to increased demand---during weekends, during the summer tourist season, or during the holiday absences of local competitors. Bakers explained that pricing otherwise would be unfair, and hence would anger and drive away customers.

\section{Monopoly Model}\label{s:monopoly}
 
We extend a simple model of monopoly pricing to include fairness concerns and subproportional inference, along the lines described in Section~\ref{s:evidence}. In this extended model, the markup charged by the monopoly is lower. Furthermore, the markup responds to marginal-cost shocks, generating some price rigidity: prices are not fixed, but they respond less than one-for-one to marginal costs.

\subsection{Assumptions}

A monopoly sells a good to a representative customer. The monopoly cannot price-discriminate, so each unit of good sells at the same price $P$. The customer cares about fairness and appraises transactional fairness by assessing the markup charged by the monopoly. Since the customer does not observe the marginal cost of production, she needs to infer it from the price. We assume that the marginal cost perceived at price $P$ is given by a belief function $C^{p}(P)$. For simplicity, we restrict $C^p(P)$ to be deterministic. Having inferred the marginal cost, the customer deduces that the markup charged by the monopoly is 
\begin{equation*}
M^{p}(P)=\frac{P}{C^{p}(P)}.
\end{equation*}
The perceived markup determines the fairness of the transaction through a fairness function $F(M^p)>0$. Both functions $C^{p}(P)$ and $F(M^p)$ are assumed to be twice differentiable.

A customer who buys the quantity $Y$ of the good at price $P$ experiences the fairness-adjusted consumption 
\begin{equation*}
Z= F(M^{p}(P)) \cdot Y.
\end{equation*}
The customer also faces a budget constraint:
\begin{equation*}
P \cdot Y + B = W,
\end{equation*}
where $W>0$ designates initial wealth, and $B$ designates remaining money balances. Fairness-adjusted consumption and money balances enter a quasilinear utility function
\begin{equation*}
\frac{\e}{\e-1} \cdot Z^{(\e-1)/\e} + B,
\end{equation*}
where the parameter $\e>1$ governs the concavity of the utility function. Given fairness factor $F$ and price $P$, the customer chooses purchases $Y$ and money balances $B$ to maximize utility subject to the budget constraint.

Finally, the monopoly has constant marginal cost $C>0$. It chooses price $P$ and output $Y$ to maximize profits $(P - C) \cdot Y$ subject to customers' demand for its good.

\subsection{Demand and Pricing}

We begin by determining customers' demand for the monopoly good. The customer chooses purchases $Y$ to maximize utility
\begin{equation*}
\frac{\e}{\e-1}  \bp{F \cdot Y}^{(\e-1)/\e} + W - P \cdot Y. 
\end{equation*}
The maximum of the customer's utility function is given by the following first-order condition:
\begin{equation*}
F^{(\e-1)/\e} \cdot Y^{-1/\e} = P.
\end{equation*}
This first-order condition yields the demand curve
\begin{equation}
Y^{d}(P)= P^{-\e} \cdot F(M^{p}(P))^{\e-1}.
\label{e:yd}\end{equation}
The price affects demand through two channels: the typical substitution effect, captured by $P^{-\e}$; and the fairness channel, captured by $F(M^{p}(P))^{\e-1}$. The fairness channel appears because the price influences the perceived markup and thus the fairness of the transaction; this in turn affects the marginal utility of consumption and demand.

We turn to the monopoly's pricing. The monopoly chooses price $P$ to maximize profits $(P-C) \cdot Y^{d}(P)$. The first-order condition is
\begin{equation*}
Y^d(P)+\bp{P-C} \od{Y^{d}}{P}=0.
\end{equation*}
We introduce the price elasticity of demand, normalized to be positive:
\begin{equation*}
E = -\odl{Y^{d}}{P} = - \frac{P}{Y^d} \cdot \od{Y^{d}}{P}.
\end{equation*}
The first-order condition then yields the classical result that
\begin{equation*} 
P= \frac{E}{E-1} \cdot C;
\end{equation*}
that is, the monopoly optimally sets its price at a markup $M=E/(E-1)$ over marginal cost.\footnote{In Appendix~A, we use the assumptions on the belief and fairness functions introduced in the next sections to verify that the first-order condition always gives the maximum of the monopoly's profit function.}

To learn more about the monopoly's markup, we compute the elasticity $E$. Using \eqref{e:yd}, we find
\begin{equation}
E= \e + (\e-1) \cdot \f \cdot \bs{1-\odl{C^{p}}{P}},
\label{e:e}\end{equation}
where $\f = -\odlx{F}{M^p}$ is the elasticity of the fairness function with respect to the perceived markup, normalized to be positive. The first term, $\e$, describes the standard substitution effect. The second term, $(\e-1) \cdot\f\cdot \bs{1-\odlx{C^{p}}{P}}$, represents the fairness channel and splits into two subterms. The first subterm, $(\e-1)\cdot \f $, appears because a higher price mechanically raises the perceived markup and thus reduces fairness. The second subterm, $- (\e-1)\cdot \f\cdot \bs{\odlx{C^{p}}{P}}$, appears because a higher price conveys information about the marginal cost and thus influences perceived markup and fairness. We now use \eqref{e:e} to compute the markup in various situations.

\subsection{No Fairness Concerns}\label{s:nofairness}

Before studying the more realistic case with fairness concerns, we examine the benchmark case in which customers do not care about fairness. 

\begin{definition} Customers who do not care about fairness have a fairness function $F(M^p)=1$.\end{definition}

Without fairness concerns, the fairness function is constant, so its elasticity is $\f=0$. According to \eqref{e:e}, the price elasticity of demand is constant: $E=\e$. This implies that the optimal markup for the monopoly takes a standard value of $\e/(\e-1)$.

Since the markup is independent of costs, changes in marginal cost are fully passed through into the price; that is, prices are flexible. Formally, the cost passthrough is
\begin{equation*}
\b = \odl{P}{C},
\end{equation*}
which measures the percentage change in price when the marginal cost increases by 1\%. The passthrough takes the value of one because
\begin{equation*}
P = \frac{\e}{\e-1} \cdot C. 
\end{equation*}

The following lemma summarizes the results:

\begin{lemma}\label{l:nofairness} When customers do not care about fairness, the monopoly sets the markup to $M=\e/(\e-1)$, and the cost passthrough is $\b=1$.\end{lemma}

\subsection{Fairness Concerns and Observable Costs}\label{s:observable}

We now introduce fairness concerns. As a preliminary step to the analysis with unobservable costs, we explore pricing when costs are observable.

To describe fairness concerns, we impose some structure on the fairness function.

\begin{definition}\label{d:fairness} Customers who care about fairness have a fairness function $F(M^p)$ that is positive, strictly decreasing, and weakly concave on $[0,M^h]$, where $F(M^h)=0$ and $M^h>\e/(\e-1)$.\end{definition}

The assumption that the fairness function strictly decreases in the perceived markup captures the pattern that customers find higher markups less fair and resent unfair transactions. The assumption that the fairness function is weakly concave means that an increase in perceived markup causes a utility loss of equal magnitude (if $F$ is linear) or of greater magnitude (if $F$ is strictly concave) than the utility gain caused by an equal-sized decrease in perceived markup. We could not find evidence on this assumption, but it seems natural that people are at least as outraged over a price increase as they are happy about a price decrease of the same magnitude. 

The properties in Definition~\ref{d:fairness} lead to the following results:

\begin{lemma}\label{l:phi} When customers care about fairness, the elasticity of the fairness function
\begin{equation*}
\f(M^p) = -\odl{F}{M^p}
\end{equation*}
is strictly positive and strictly increasing on $(0,M^h)$, with $\lim_{M^p\to 0} \f(M^p) = 0$ and $\lim_{M^p\to M^h} \f(M^p) = +\infty$. As an implication, the superelasticity of the fairness function
\begin{equation*}
\s = \odl{\f}{M^p}
\end{equation*}
is strictly positive on $(0,M^h)$.\end{lemma}   

The proof is simple algebra and relegated to Appendix~A. The property that the superelasticity of the fairness function is positive plays a central role in the analysis. It means that the fairness function is more elastic at higher perceived markups. This property follows from Definition~\ref{d:fairness} because a positive, decreasing, and weakly concave function always has positive superelasticity.\footnote{The concavity of the fairness function is not a necessary condition for the results in the paper: the necessary condition is that the superelasticity of the fairness function is positive. This occurs with concave functions but also with other not-too-convex functions. For example, the logistic function $F(M^{p})=1/[1+(M^{p})^{\t}]$ with $\t>0$ is not concave but it has a positive superelasticity: $\s = \t/[1+(M^{p})^{\t}]>0$. All the results would carry over with a logistic fairness function. We limit ourselves to concave fairness functions instead of allowing for any fairness function with a positive superelasticity because we find such restriction more natural and easier to interpret.}

Since the marginal cost is assumed to be observable, customers correctly perceive marginal cost ($C^p = C$), so the perceived markup equals the true markup ($M^{p}=M$). From \eqref{e:e}, we see that the price elasticity of demand is $E=\e+(\e-1) \cdot \f(M)>\e$; therefore, the markup charged by the monopoly satisfies
\begin{equation}
M=1+\frac{1}{\e-1}\cdot \frac{1}{1+\f(M)}.
\label{e:mo}\end{equation}
Since $\f(M)$ is strictly increasing from $0$ to $+\infty$ when $M$ increases from $0$ to $M^h$ (Lemma~\ref{l:phi}), the right-hand side of the equation is strictly decreasing from $\e/(\e-1)$ to 1 when $M$ increases from $0$ to $M^h > \e/(\e-1)>1$. We infer that the fixed-point equation \eqref{e:mo} admits a unique solution, located between 1 and $\e/(\e-1$). Therefore, the markup $M$ is well-defined and $M\in (1,\e/(\e-1))$.

The next lemma records the results:

\begin{lemma}\label{l:observable} When customers care about fairness and observe costs, the monopoly's markup $M$ is implicitly defined by \eqref{e:mo}. This implies that $M \in (1,\e/(\e-1))$ and the cost passthrough is $\b=1$. Hence, the markup is lower than without fairness concerns, but the cost passthrough is identical.\end{lemma}

Without fairness concerns, the price affects demand solely through customers' budget sets. With fairness concerns and observable marginal costs, the price also influences the perceived fairness of the transaction: when the price is high relative to marginal cost, customers deem the transaction to be less fair, which reduces the marginal utility from consuming the good and hence demand. Consequently, the monopoly's demand is more price elastic than without fairness concerns, which forces the monopoly to charge a lower markup.

However, \eqref{e:mo} shows that with fairness concerns and observable costs, the markup does not depend on costs, as in the absence of fairness concerns. Since changes in marginal cost do not affect the markup, they are completely passed through into price: prices remain flexible.

\subsection{Fairness Concerns and Rational Inference of Costs}\label{s:rational}

Next, we combine fairness concerns with unobservable marginal costs, beginning with a final preliminary case in which customers rationally invert the price to uncover the hidden marginal cost. In this case, the model takes the form of a simple signaling game in which the monopoly learns its marginal cost and chooses a price, before customers observe the monopoly's price---but not its marginal cost---and formulate demand. 

Let $[0, C^h]$ be the set of all possible marginal costs for the monopoly. The monopoly knows its marginal cost $C \in [0, C^h]$, but customers do not; instead, customers have non-atomistic prior beliefs over $[0, C^h]$.

A pure-strategy perfect Bayesian equilibrium (PBE) of this game comprises three elements: a pure strategy for the monopolist, which is a mapping $P: [0, C^h] \to \R_+$ that selects a price for every possible value of marginal cost; a belief function for customers, which is a mapping $C^p: \R_+ \to [0, C^h]$ that determines a marginal cost for every possible price; and a pure strategy for customers, which is a mapping $Y^d: \R_+ \to \R_+$ that selects a quantity purchased for every possible price.\footnote{Strictly speaking, $C^p$ should allow the consumer to hold probabilistic beliefs about the firm's marginal cost given price, but we sidestep this subtlety because it does not affect our analysis.} 

We look for a PBE that is fully separating: the monopoly chooses different prices for different marginal costs, which allows a rational customer who knows the monopoly's equilibrium strategy and observes the price to deduce marginal cost. We claim the existence of a PBE in which the monopolist uses the strategy $P(C)= \e \cdot C/(\e-1)$; customers believe $C^p(P) = (\e-1) P/\e$ if $P\in \Pc \equiv \bs{0,\e C^h/(\e-1)}$, and $C^p(P) = 0$ otherwise; and customers demand $Y^{d}(P)= P^{-\e} \cdot F(P/C^p(P))^{\e-1}$. In such a PBE, customers correctly infer marginal costs from prices on the equilibrium path ($P\in \Pc$) and infer the worst from prices off the equilibrium path ($P\notin \Pc$), namely that the firm has zero marginal cost and infinitely high markup.

The argument proceeds in three steps. First, given their beliefs, customers' demand is indeed optimal, as we have shown in \eqref{e:yd}. Second, given the monopolist's strategy, customers' beliefs are indeed correct for any equilibrium price. Third, given customers' beliefs and demand, the monopolist's strategy is optimal. Indeed, given customers' beliefs for $P\in \Pc$, we have $\odlx{C^{p}}{P}=1$. Then, according to \eqref{e:e} (which is implied by customers' strategy), the price elasticity of demand for any price on $\Pc$ is $E=\e$. Hence, the monopolist optimally charges $P= \e C/(\e-1)$. Finally, the monopoly has no incentive to charge some price not belonging to $\Pc$, which would lead customers to perceive an infinite markup, bringing the fairness factor, demand, and profits to zero. 

The following lemma records the findings:

\begin{lemma}\label{l:rational} When customers care about fairness and rationally infer costs, there is a PBE in which the monopoly uses the markup $M=\e/(\e-1)$, and customers learn marginal cost from price. In this PBE, the cost passthrough is $\b=1$. Hence, in this PBE, the markup and cost passthrough are the same as without fairness concerns.\end{lemma}
 
The lemma shows that when customers care about fairness and rationally infer costs, there is a PBE in which fairness does not play a role. With fairness concerns, the price affects demand not only by changing customers' budget sets but also by changing the perceived markup. In this equilibrium, however, after observing any price chosen by the monopoly, customers perceive the same markup $\e/(\e-1)$. The second channel through which price could affect demand closes, so the monopoly sets the standard markup $\e/(\e-1)$. Since the markup does not depend on marginal cost, changes in marginal cost are fully passed through into prices: prices are flexible again.

Of course, there may exist other equilibria beside the one described in Lemma~\ref{l:rational}. A pooling PBE may exist in which all types of the firm charge the same price $P>C^h$, and consumers believe that a firm who prices otherwise has zero marginal cost. However, this and other non-fully-separating PBEs fail standard signaling refinements.\footnote{Only a separating PBE satisfies the D1 Criterion from \ct{CK87}. Intuitively, consumers ought to interpret a price $P'>P$ as coming from type $C=C^h$ rather than type $C=0$, which undermines the pooling equilibrium. Indeed, if consumers demand no less at $P'$ than in equilibrium, then all types of firm benefit from deviating; if consumers demand less at $P'$ than in equilibrium, then the highest-cost firm strictly benefits whenever any other type of firm weakly benefits. On these grounds, the D1 Criterion suggests that consumers should interpret $P'>P$ as coming from the highest marginal-cost firm.} Because the linear PBE in Lemma~\ref{l:rational} is so simple and robust, it is more plausible than any alternative, which suggests that fairness is unlikely to matter when customers rationally infer costs.
 
\subsection{Fairness Concerns and Subproportional Inference of Costs}\label{s:subproportional}

We turn to the main case of interest: customers care about fairness and subproportionally infer costs from prices. In this case, the fairness function satisfies Definition~\ref{d:fairness}, and the belief function takes the following form:

\begin{definition}\label{d:subproportional} Customers who update subproportionally use the belief-updating rule
\begin{equation}
C^{p}(P)=\bp{C^{b}}^{\g}  \bp{\frac{\e-1}{\e} P}^{1-\g},
\label{e:cp}\end{equation}
where $C^b > (\e-1) \cdot (M^h)^{-1/\g} \cdot C/\e$ is a prior point belief about marginal cost, and $\g \in(0,1]$ governs the extent to which beliefs anchor on that prior belief.\end{definition}

We have seen evidence that people do not sufficiently introspect about the relationship between price and marginal cost, which leads them to underinfer the information conveyed by the price, and that they tend to think proportionally. The inference rule \eqref{e:cp} geometrically averages no inference with proportional inference, so it encompasses these two types of error.

First, customers underinfer marginal costs from price by clinging to their prior belief $C^b$. The parameter $\g \in (0,1]$ measures the degree of underinference. When $\g=1$, customers do not update at all about marginal cost based on price; they naively maintain their prior belief $C^{b}$, irrespective of the price they observe. When $\g\in(0,1)$, customers do infer something from the price, but not enough. 

Moreover, insofar as they infer something, they infer that marginal cost is proportional to price, given by $(\e-1) P/\e$. Such proportional inference represents a second error: underinference pertains to how much customers infer, whereas proportional inference describes what customers infer in as much as they do infer. The updating rule has the property that in the limit as $\g=0$, customers infer rationally. Indeed, when $\g=0$, the monopoly optimally sets the markup $\e/(\e-1)$, which makes $(\e-1) P/\e$ the marginal cost at price $P$, and proportional inference agrees with rational inference. When $\g \in (0,1)$, however, the monopoly does not find it optimal to mark up proportionally, and proportional inference becomes an error.

Last, we impose a constraint on the parameter $C^b$ such that the perceived markup falls below $M^h$ when the firm prices at marginal cost; this is necessary for equilibrium existence.

Despite its apparent arbitrary nature, the assumption of subproportional inference has close ties to game-theoretic models of failure of contingent thinking. It is related to the concept of cursed equilibrium, developed by \ct{ER05}, and to the concept of analogy-based-expectation equilibrium, developed by \ct{J05} and extended to Bayesian games by \ct{JK08}. Both concepts propose mechanisms that can be used to explain why people might fail to account for the information that equilibrium prices reveal about marginal costs.\footnote{In fact, with $\g=1$, the beliefs given by \eqref{e:cp} resemble those in a fully cursed equilibrium and the coarsest analogy-based-expectation equilibrium, when recasting our model as a Bayesian game, as in Section~\ref{s:rational}. In these equilibrium concepts, an unsophisticated household infers nothing about marginal cost from any economic variable. Consequently, a consumer with average prior beliefs about marginal cost equal to $C^b$ would continue to perceive marginal costs with mean $C^b$ given any price.} Subproportional inference is also related to the cursed-expectation equilibrium developed by \ct{ERV19} as an alternative to rational-expectations equilibrium in markets.\footnote{In a cursed-expectation equilibrium of a model in which traders endowed with private information trade a risky asset, each trader forms an expectation about the value of the asset equal to a geometric average of her expectation conditional upon her private signal alone and her expectation conditional upon both her private signal and the market price. Traders' expectations therefore take the form of a weighted average of naive beliefs and correct beliefs. The two rules differ in that consumers in our model average naive beliefs with a particular form of incorrect beliefs (proportional inference); to include rational updating as a limit case, we calibrate the updating rule to match correct equilibrium beliefs for the case in which all consumers are rational. We adopt this approach for its tractability.}

Subproportional inference also draws upon several well-documented psychological biases. Customers in our model are coarse thinkers in the sense of \ct{MSS08} because they do not distinguish between scenarios where price changes reflect changes in cost and those where they reflect changes in markup. The underinference could also be a form of the anchoring heuristic documented by \ct{TK74}: consumers understand that higher prices reflect higher marginal costs but do not adjust sufficiently their estimate of the marginal cost. Altogether, the updating rule \eqref{e:cp} captures the well-known bias that people do not update their beliefs sufficiently from available information.

\paragraph{Analytical Results}  Plugging the belief-updating rule \eqref{e:cp} into $M^p= P/C^p$ gives the following:

\begin{lemma}\label{l:subproportional} When customers update subproportionally, they perceive the monopoly's markup to be
\begin{equation*}
M^{p}(P)=\bp{\frac{\e}{\e-1}}^{1-\g}  \bp{\frac{P}{C^{b}}}^{\g},
\end{equation*}
which is a strictly increasing function of the observed price $P$.\end{lemma}

Customers appreciate that higher prices signal higher marginal costs. But by underappreciating the strength of the relationship between price and marginal cost, customers partially misattribute higher prices to higher markups. Consequently, they regard higher prices as less fair. 

As the belief function $M^{p}(P)$ and fairness function $F(M^{p})$ are differentiable, customers enjoy an infinitesimal price reduction as much as they dislike an infinitesimal price increase. Therefore, the monopoly's demand curve \eqref{e:yd} has no kinks, unlike in pricing theories based on loss aversion \cp{HK08}.

Combining \eqref{e:e} and \eqref{e:cp}, we then find that the price elasticity of demand satisfies
\begin{equation}
E=\e+(\e-1) \cdot \g  \cdot \f(M^p).
\label{e:e2}\end{equation}
We have seen that without fairness concerns ($\f=0$), or with rational inference ($\g=0$), the price elasticity of demand is constant, equal to $\e$. That result changes here. Since $\g>0$, the price elasticity of demand is always greater than $\e$. Moreover, since $\f(M^{p})$ is increasing in $M^{p}$ and $M^{p}(P)$ in $P$, the price elasticity of demand is increasing in $P$. These properties have implications for the markup charged by the monopoly, $M=E/(1-E)$.

\begin{proposition}\label{p:subproportional} When customers care about fairness and update subproportionally, the monopoly's markup $M$ is implicitly defined by
\begin{equation}
M=1+\frac{1}{\e-1}\cdot \frac{1}{1+\g \f(M^p(M \cdot C))},
\label{e:m}\end{equation}
which implies that $M \in (1,\e/(\e-1))$. Furthermore, the cost passthrough is given by
\begin{equation*}
\b=1\bigg/\bc{1+\frac{\g^{2}\f \s}{\bp{1+\g \f} \bs{\e+(\e-1) \g\f}}},
\end{equation*} 
which implies that $\b\in(0,1)$. Hence, the markup is lower than without fairness concerns or with rational inference; and unlike without fairness concerns or with rational inference, the cost passthrough is incomplete.\end{proposition}

The proof is relegated to Appendix~A, but the intuition is simple. First, when customers care about fairness but underinfer marginal costs, they become more price-sensitive. Indeed, an increase in the price increases the opportunity cost of consumption---as in the case without fairness---and also increases the perceived markup, which reduces the marginal utility of consumption and therefore demand. This heightened price-sensitivity raises the price elasticity of demand above $\e$ and pushes the markup below $\e/(\e-1)$. 

Second, after an increase in marginal cost, the monopoly optimally lowers its markup. This occurs because customers underappreciate the increase in marginal cost that accompanies a higher price. Since the perceived markup increases, the price elasticity of demand increases. In response, the monopoly reduces its markup, which mitigates the price increase. Thus, our model generates incomplete passthrough of marginal cost into price---a mild form of price rigidity. Furthermore, customers err in believing that transactions are less fair when the marginal cost increases: transactions actually become more fair.

\paragraph{Comparison with Microevidence}  The result that prices do not fully respond to marginal-cost shocks accords well with evidence on real firm behavior. First, using matched data on product prices and producers' unit labor cost in Sweden, \ct{CS12} find a passthrough of idiosyncratic marginal-cost changes into prices of only $0.3$. Second, using production data for Indian manufacturing firms, \ct[Table~7]{DGK16} find that following trade liberalization in India, marginal costs fell significantly due to the import tariff reduction, yet prices failed to fall in step: they estimate passthroughs between $0.3$ and $0.4$. Third, using production and cost data for Mexican manufacturing firms, \ct[Table~7]{CCW17} also find a modest passthrough of idiosyncratic marginal-cost changes into prices: between $0.2$ and $0.4$. Last, combining production data for US manufacturing firms with data on energy prices and consumption, \ct[Tables~5 and 6]{GSW19} find a moderate passthrough of marginal-cost changes caused by energy-price variations into prices: between $0.5$ and $0.7$. Taking the midpoint estimates from the four studies, we find an average passthrough of $0.3+0.35+0.3+0.6=0.4$. Such cost passthrough is well below~1. 

Additionally, our theory predicts that when customers care about fairness, the passthrough of marginal costs into prices is markedly different when costs are observable and when they are not. The passthrough is one when costs are observable (Lemma~\ref{l:observable}) but is strictly below one when costs are not observable (Proposition~\ref{p:subproportional}). \ct{KaLS91}, \ct{KLS91}, and \ct{RT04} provide experimental evidence consistent with this result: they find that after a cost shock, prices adjust more when costs are observable than when they are not.

\paragraph{Comparison with the Literature}  Price rigidity in our model arises from a nonconstant price elasticity of demand, which creates variation in markups after cost shocks. Other models share the feature that variable price elasticity leads to price rigidity. In international economics, these models have been used to explain the behavior of exchange rates and prices \cp{D87,BeF01,AB08}. In macroeconomics, they have been used to create real rigidities that amplify nominal rigidities \cp{K95,DK05,EF07}. Unlike many of these models, our model does not make reduced-form assumptions about the utility function or demand curve to generate a nonconstant price elasticity of demand but instead provides a microfoundation. 

\paragraph{Additional Analytical Results}  To obtain further results, we introduce a simple fairness function that satisfies all the requirements from Definition~\ref{d:fairness}:
\begin{equation}
F(M^p) = 1- \t \cdot \bp{M^p - \frac{\e}{\e-1}},
\label{e:f}\end{equation}
where $\t>0$ governs the intensity of fairness concerns. A higher $\t$ means that a consumer grows more upset when consuming an overpriced item and more content when consuming an underpriced item. The fairness function reaches $1$ when the perceived markup equals $\e/(\e-1)$; then fairness-adjusted consumption coincides with actual consumption. When the perceived markup exceeds $\e/(\e-1)$, the fairness function falls below one; and when the perceived markup lies below $\e/(\e-1)$, the fairness function surpasses one.

Furthermore, to compare different industries or economies, we focus on a situation in which customers have acclimated to prices by coming to judge firms' markups as acceptable: $C^b$ adjusts so $M^p = \e/(\e-1)$ and $F=1$. Acclimation is likely to occur eventually within any industry or economy, once customers have faced the same prices for a long time.\footnote{As noted by \ct[p.~730]{KKT86}, ``Psychological studies of adaption suggest that any stable state of affairs tends to become accepted eventually, at least in the sense that alternatives to it no longer come to mind. Terms of exchange that are initially seen as unfair may in time acquire the status of a reference transaction\ldots. [People] adapt their views of fairness to the norms of actual behavior.'' The belief-updating rule~\eqref{e:cpj} introduced in the New Keynesian model has the property that for any initial belief, people eventually become acclimated.} 

We then obtain the following comparative statics:

\begin{corollary}\label{c:subproportional} Assume that customers care about fairness according to the fairness function \eqref{e:f}, infer subproportionally, and are acclimated. Then the monopoly's markup is given by 
\begin{equation*}
M = 1+\frac{1}{\bp{1+\g\t} \e-1}.
\end{equation*} 
The markup decreases with the competitiveness of the market ($\e$), concern for fairness ($\t$), and degree of underinference ($\g$). And the cost passthrough is given by 
 \begin{equation*}
\b = 1\bigg/\bc{1+\frac{\g^{2} \t \bs{\bp{1+\t} \e-1}}{ \bp{\e-1}\bp{1+\g\t}\bs{\bp{1+\g\t} \e-1}}}.
\end{equation*} 
The passthrough increases with the competitiveness of the market ($\e$); it decreases with the concern for fairness ($\t$) and degree of underinference ($\g$).\end{corollary}

The proof is in Appendix~A; it applies Proposition~\ref{p:subproportional} to the fairness function~\eqref{e:f} under acclimation.

\paragraph{Comparison with Additional Microevidence}  Our theory predicts that the cost passthrough is higher in more-competitive markets. This property echoes the finding by \ct{C86} that prices are less rigid in less-concentrated industries. It is also consistent with the finding by \ct{AIK14} that firms with higher market power have a lower passthrough of cost shocks driven by exchange-rate fluctuations.

Our theory also predicts that the passthrough is lower---so prices are more rigid---in markets that are more fairness-oriented. This property could contribute to explain the finding by \ct{K07} that retail prices were more rigid in 1889--1891 than in 1997--1999. \name{K07} emphasizes that the relationship between retailers and customers was much more personal in the 19th century than today.\footnote{\name{K07} notes: ``In 1889--1891 retailing often occurred in small one- or two-person shops, retailers supplied credit to the customers, and retailers usually delivered the purchases to the customer's home at no extra charge. Today retailing occurs in large stores, a third party supplies credit, and the customer takes his own items home. These changes decrease both the business and personal relationship between the retailer and the customer'' (p.~2008).} This more personal relationship could have made the retail sector more fairness-oriented, which would help explain, according to our theory, its greater historical price rigidity. This channel could also help explain the finding by \ct[Table~8]{NS07} that prices are more rigid in the service sector than elsewhere, since relationships between buyers and sellers are more personal in the service industry.

\section{New Keynesian Model}\label{s:nk}

We now explore the macroeconomic implications of the pricing theory developed in Section~\ref{s:monopoly}. To that end, we embed it into a New Keynesian model as a substitute for usual pricing frictions---either \ct{C83} pricing or \ct{R82} pricing. We find that when customers care about fairness and infer subproportionally, the price markup depends on the rate of inflation; thus, monetary policy is nonneutral in both short and long run. (Derivations are relegated to Appendix~B.)

\subsection{Assumptions}

A continuum of firms indexed by~$j\in[0,1]$ and a continuum of households indexed by $k\in [0,1]$ make up the economy. Firms use labor services to produce goods. Households supply labor services, consume goods, and save using riskless nominal bonds. Since goods are imperfect substitutes for one another, and labor services are also imperfect substitutes, firms exercise some monopoly power on the goods market, and households exercise some monopoly power on the labor market.

\paragraph{Fairness Concerns}  Households cannot observe firms' marginal costs. When a household purchases good~$j$ at price $P_{j}(t)$ in period~$t$, it infers that firm~$j$'s marginal cost is $C^{p}_{j}(t)$. The model is dynamic so it provides a natural candidate for the anchor that households use when inferring costs: last period's perception of marginal cost. Hence, instead of being given by \eqref{e:cp} as in the monopoly model, households' perception of firm~$j$'s marginal cost at time~$t$ is given by
\begin{equation}
C^{p}_{j}(t)= \bs{C^{p}_{j}(t-1)}^{\g}\bs{\frac{\e-1}{\e} P_{j}(t)}^{1-\g},
\label{e:cpj}\end{equation}
where $C^{p}_{j}(t-1)$ is last period's perceived cost, and $\g\in(0,1)$ is the degree of underinference.

Having inferred the marginal cost, households deduce that the markup charged by firm~$j$ is $M^{p}_{j}(t)=P_{j}(t)/C^{p}_{j}(t)$. This perceived markup determines the fairness of the transaction with firm~$j$, measured by $F_{j}(M^{p}_{j}(t))$. The fairness function $F_{j}$, specific to good~$j$, satisfies the conditions listed in Definition~\ref{d:fairness}. The elasticity of $F_{j}$ with respect to $M^{p}_{j}$ is $\f_{j} = -\odlx{F_{j}}{M^{p}_{j}}$.

An amount $Y_{jk}(t)$ of good~$j$ bought by household~$k$ at a unit price $P_{j}(t)$ yields a fairness-adjusted consumption 
\begin{equation*}
Z_{jk}(t)= F_{j}(M^{p}_{j}(P_{j}(t))) \cdot Y_{jk}(t). 
\end{equation*}
Household~$k$'s fairness-adjusted consumption of the various goods aggregates into a consumption index 
\begin{equation*}
Z_{k}(t) = \bs{\int_{0}^{1} Z_{jk}(t)^{(\e-1)/\e}\,dj}^{\e/(\e-1)},
\end{equation*}
where $\e>1$ is the elasticity of substitution between different goods. The price of one unit of the consumption index at time~$t$ is given by the price index
\begin{equation*}
X(t)=\bc{\int_0^1\bs{\frac{P_{j}(t)}{F_{j}(M^{p}_{j}(P_{j}(t)))}}^{1-\e}\,dj}^{1/(1-\e)}.
\end{equation*}

\paragraph{Households}  Household~$k$ derives utility from consuming goods and disutility from working. Its expected lifetime utility is
\begin{equation*}
\E[0] \sum_{t=0}^{\infty}\d^{t} \bs{\ln(Z_{k}(t))-\frac{N_{k}(t)^{1+\eta}}{1+\eta}},
\end{equation*}
where $\E[t]$ is the mathematical expectation conditional on time-$t$ information,  $\d\in(0,1)$ is the discount factor, $N_{k}(t)$ is its labor supply, and $\eta > 0$ is the inverse of the Frisch elasticity of labor supply. 

To smooth consumption over time, households trade one-period bonds. In period~$t$, household~$k$ holds $B_{k}(t)$ bonds. Bonds purchased in period~$t$ have a price $Q(t)$, mature in period~$t+1$, and pay one unit of money at maturity.

Household~$k$'s consumption-savings decisions in each period~$t$ must obey the constraint
\begin{equation*}
\int_{0}^{1} P_{j}(t)  Y_{jk}(t)\,dj + Q(t)  B_{k}(t) = W_{k}(t)  N_{k}(t)+ B_{k}(t-1)+V_{k}(t),
\end{equation*}
where $W_{k}(t)$ is the wage rate for labor service~$k$, and $V_{k}(t)$ are dividends from firm ownership. In addition, household~$k$ satisfies a solvency constraint that prevents Ponzi schemes.

Finally, in each period~$t$, household~$k$ chooses purchases $Y_{jk}(t)$ for each $j\in[0,1]$, labor supply $N_{k}(t)$, bond holdings $B_{k}(t)$, and wage rate $W_{k}(t)$. The household's objective is to maximize its expected utility subject to the budget constraint, to the solvency constraint, and to firms' demand for labor service~$k$. The household takes as given its initial endowment of bonds $B_{k}(-1)$, all fairness factors $F_{j}(t)$, all prices $P_{j}(t)$ and $Q(t)$, and dividends $V_{k}(t)$.
 
\paragraph{Firms}  Firm~$j$ hires labor to produce output using the production function
\begin{equation}
Y_{j}(t)=A_{j}(t)  N_{j}(t)^{\a},
\label{e:yj}\end{equation}
where $Y_{j}(t)$ is output of good~$j$, $A_{j}(t)>0$ is its technology level, $\a\in(0,1]$ is the extent of diminishing marginal returns to labor, and
\begin{equation*}
N_{j}(t)= \bs{\int_{0}^1 N_{jk}(t)^{(\n-1)/\n}\,dk}^{\n/(\n-1)}
\end{equation*}
is an employment index. In the index, $N_{jk}(t)$ is the quantity of labor service~$k$ hired by firm~$j$, and $\n>1$ is the elasticity of substitution between different labor services. The technology level $A_{j}(t)$ is stochastic and unobservable to households---making the firm's marginal cost unobservable.

Each period~$t$, firm~$j$ chooses output $Y_{j}(t)$, price $P_{j}(t)$, and employment levels $N_{jk}(t)$ for all $k\in[0,1]$. The firm's objective is to maximize the expected present-discounted value of profits
\begin{equation*}
\E[0]\sum_{t=0}^{\infty}\G(t) \bs{P_{j}(t) Y_{j}(t)-\int_{0}^{1}W_{k}(t)  N_{jk}(t)\,dk},
\end{equation*}
where $\G(t)$ is the stochastic discount factor for period-$t$ nominal payoffs, subject to the production constraint~\eqref{e:yj}, to demand for good~$j$, and to the law of motion of the perceived marginal cost \eqref{e:cpj}. The firm takes as given the initial belief about its marginal cost $C^p_{j}(-1)$, all wage rates $W_{k}(t)$, and discount factors $\G(t)$. Its profits accrue to households as dividends.

\paragraph{Monetary Policy}  The nominal interest rate is determined by a simple monetary-policy rule:
\begin{equation}
i(t)=i_{0}(t)+\p \pi(t), 
\label{e:taylor}\end{equation}
where $i_{0}(t)$ is a stochastic exogenous component, $\pi(t)$ is the inflation rate, and $\p>1$ governs the response of the interest rate to inflation.

\paragraph{Symmetry}  We assume a symmetric economy. All households receive the same bond endowment $B(-1)$ and same dividends $V(t)$. All firms share a common technology $A(t)$, face the same fairness function $F$, and are believed to have the same initial cost $C^p(-1)$. Hence, all households behave identically, as do all firms.

\paragraph{Notation}  Since the equilibrium is symmetric, we drop subscripts~$j$ and~$k$ to denote the equilibrium values taken by the variables. We also denote the steady-state value of any variable $H(t)$ by $\ol{H}$. And for any variable $H(t)$ except the inflation and interest rates, we denote the logarithmic deviation from steady state by $\wh{h}(t)\equiv \ln(H(t))-\ln(\ol{H})$. For the inflation and interest rates, we denote the deviation from steady state by $\wh{\pi}(t)\equiv \pi(t)-\ol{\pi}$, $\wh{i_0}(t) \equiv i_0(t)-\ol{i_0}$, and $\wh{i}(t) \equiv i(t) - \ol{i}$.

\subsection{Demand for Goods and Pricing}

Households and firms behave exactly as in the textbook New Keynesian model, except that fairness concerns modify consumers' demand and, consequently, firms' pricing.

The demand for good~$j$ from all households is
\begin{equation*}
Y^{d}\of{t,P_{j}(t),C_{j}^{p}(t-1)} = Z(t) \bs{\frac{P_{j}(t)}{X(t)}}^{-\e} F\of{\bp{\frac{\e}{\e-1}}^{1-\g}\bs{\frac{P_{j}(t)}{C_{j}^{p}(t-1)}}^{\g}}^{\e-1},
\end{equation*}
where $Z(t) = \int_0^1 Z_{k}(t)\,dk$ describes the level of aggregate demand. The price of good~$j$ appears twice in the demand curve: as part of the relative price $P_j/X$; and as part of the fairness factor $F$. This second element leads to unconventional pricing. 

As in the monopoly model, fairness affects pricing through the price elasticity of demand $E$, which satisfies~\eqref{e:e2}. Unlike in the monopoly model, however, the profit-maximizing markup is not $E/(E-1)$ because $E$ does not capture the effect of the current price on future beliefs and thus future demands. Instead, in equilibrium, firms set their price markup $M$ such that
\begin{equation}
\frac{M(t)-1}{M(t)} E(t) = 1-\d\g+\d\E[t]{\frac{M(t+1)-1}{M(t+1)} \bs{E(t+1)-(1-\g)\e}}.
\label{e:em}\end{equation}
The gap between $M(t)$ and $E(t)/[E(t)-1]$ reflects how much today's price affects future perceived marginal costs, demands, and profits. Conversely, if firms do not care about the future ($\d=0$), the equation reduces to $M(t) = E(t)/[E(t)-1]$.

The price markup plays a critical role because it directly determines employment:
\begin{equation}
N(t)=\bs{\frac{(\n-1) \a }{\n} \cdot  \frac{1}{M(t)}}^{1/(1+\eta)}.
\label{e:n}\end{equation}
Employment is strictly decreasing in the price markup because in equilibrium the price markup is the inverse of the real marginal cost, which is itself increasing in employment. Since a higher price markup implies a lower real marginal cost, it also implies lower employment.

\subsection{Calibration}\label{s:calibration}

Before simulating the model, we calibrate it to US data. To set the values of the fairness-related parameters, we use new evidence on price markups and cost passthroughs. For the other parameters, we rely on standard evidence. The calibrated values of the parameters are summarized in Table~\ref{t:calibration}.

\paragraph{Fairness Function}  We set the shape of the fairness function $F$ to \eqref{e:f}. This simple functional form has two advantages. First, it introduces only one new parameter, $\t>0$, which governs the concern for fairness. Second, it produces a fairness factor equal to one at the zero-inflation steady state. Indeed, in such steady state, the perceived price markup is $\ol{M^p}=\ol{P}/\ol{C^p}=\e/(\e-1)$, as shown by \eqref{e:cpj}, and so the fairness factor is $\ol{F}=1$. Thus, with no trend inflation, customers acclimate and are neither happy nor unhappy about markups.

\begin{table}[t]
\caption{Parameter values in simulations}
\footnotesize\begin{tabular*}{\textwidth}{@{\extracolsep{\fill}}lll}\toprule
	Value & Description & Source or target\\
	\midrule
	\multicolumn{3}{c}{A. Common parameters}\\
 	$\d=0.99$ & Quarterly discount factor & Annual rate of return $=4\%$ \\
	$\a = 1$ & Shape of production function & Labor share $=2/3$ \\
   $\eta=1.1$ & Inverse of Frisch elasticity of labor supply & \ct[Table~2]{CGM13} \\
   $\p=1.5$ & Response of nominal interest rate to inflation & \ct[p.~52]{G08} \\
   $\m^i=0.75$ & Persistence of monetary shock & \ct[p.~52]{G08}, \ct[p.~26]{G10}\\
   $\m^a=0.9$ & Persistence of technology shock & \ct[p.~55]{G08}\\
   \midrule
   \multicolumn{3}{c}{B. Parameters of the New Keynesian model with fairness}\\
   $\e=2.23$ & Elasticity of substitution across goods & Steady-state price markup $=1.5$ \\
	$\t=9$ & Fairness concern & Instantaneous cost passthrough $=0.4$ \\
   $\g=0.8$ & Degree of underinference & Two-year cost passthrough $=0.7$  \\
   \midrule
   \multicolumn{3}{c}{C. Parameters of the textbook New Keynesian model}\\
   $\e=3$ & Elasticity of substitution across goods & Steady-state price markup $=1.5$ \\
   $\x=0.67$ & Share of firms keeping price unchanged & Average price duration $=3$ quarters \\
\bottomrule\end{tabular*}
\note{The parameter values described in the table are obtained in Section~\ref{s:calibration}.}
\label{t:calibration}\end{table}

\paragraph{Fairness-Related Parameters}  We then calibrate the three parameters central to our theory: the fairness parameter $\t$, the inference parameter $\g$, and the elasticity of substitution across goods $\e$. These parameters jointly determine the average value of the price markup and its response to shocks---which determines the cost passthrough. Hence, for the calibration, we match evidence on price markups and cost passthroughs. We target three empirical moments: average price markup, short-run cost passthrough, and long-run cost passthrough.

First, using firm-level data, \ct[p.~575]{DEG20} estimate price markups in the United States. They find that the average markup across the US economy hovers between $1.2$ and $1.3$ in the 1955--1980 period, rises from $1.2$ in 1980 to $1.5$ in 2000, remains around $1.5$ until 2014, before spiking to $1.6$ in 2016. As the markup averages $1.5$ between 2000 and 2016, we adopt this value as a target.\footnote{The aggregate markup computed by \ct{DEG20} is commensurate to markups measured in specific industries and goods in the United States. In the automobile industry, \ct[p.~882]{BLP95} estimate that on average $(P-C)/P = 0.239$, which translates into a markup of $M=P/C = 1/(1-0.239) = 1.3$. In the ready-to-eat cereal industry, \ct[Table~8]{N01} finds that a median estimate of $(P-C)/P$ is $0.372$, which translates into a markup of $M=P/C = 1/(1-0.372) = 1.6$. In the coffee industry, \ct[Table~6]{NZ10} also estimate a markup of $1.6$. For most national-brand items retailed in supermarkets, \ct[p.~166]{BBD03} discover that markups range between $1.4$ and $2.1$. Finally, earlier work surveyed by \ct[pp.~260--267]{RW95} finds similar markups: between $1.2$ and $1.7$ in the industrial-organization literature, and around $2$ in the marketing literature.}

Second, in the United States, Sweden, India, and Mexico, the short-run cost passthrough is estimated between $0.2$ and $0.7$, with an average value of~$0.4$ (Section~\ref{s:subproportional}). Hence, we target a short-run cost passthrough of~$0.4$.

Third, \ct[Table~7.4]{BG14} provide estimates of the long-run exchange-rate passthrough for the United States and seven other countries. The exchange-rate passthrough measures the response of import prices to exchange-rate shocks. Its level may not reflect that of the cost passthrough, because marginal costs may not vary one-for-one with exchange rates, but there is no reason for the two passthroughs to have different dynamics \cp{AIK14}. The immediate exchange-rate passthrough is estimated at $0.4$, and the two-year exchange-rate passthrough at~$0.7$. Based on these dynamics, and the fact that the immediate cost passthrough is also~$0.4$, we target a two-year cost passthrough of~$0.7$.

We then simulate the dynamics of a firm's price in response to an unexpected and permanent increase in its marginal cost (see Appendix~B.5). We find that the fairness parameter $\t$ primarily affects the level of the cost passthrough, while the inference parameter $\g$ primarily affects its persistence. Based on the simulations, we set $\e=2.23$, $\t=9$, and $\g=0.8$. This calibration allows us to achieve a steady-state price markup of~$1.5$, an instantaneous cost passthrough of~$0.4$, and a two-year cost passthrough of~$0.7$.

\paragraph{Other Parameters}  We set the labor-supply parameter to $\eta=1.1$, which gives a Frisch elasticity of labor supply of $1/1.1=0.9$. This value is the median microestimate of the Frisch elasticity for aggregate hours \cp[Table~2]{CGM13}. We then set the quarterly discount factor to $\d=0.99$, giving an annual rate of return on bonds of~$4\%$. We set the production-function parameter to $\a=1$. This calibration guarantees that the labor share, which equals $\a/\ol{M}$ in steady state, takes its conventional value of~$2/3$. Last, we calibrate the monetary-policy parameter to $\p=1.5$, which is consistent with observed variations in the federal funds rate \cp[p.~52]{G08}.

\paragraph{Parameters of the Textbook New Keynesian Model}  We also calibrate a textbook New Keynesian model (described in Appendix~C), which we will use as a benchmark in simulations. For the parameters common to the two models, we use the same values---except for $\e$. In the textbook model, the steady-state price markup is $\e/(\e-1)$, so we set $\e=3$ to obtain a markup of~$1.5$. 

We also need to calibrate a parameter specific to the textbook model: $\x$, which governs price rigidity. To generate  price rigidity, the New Keynesian literature uses either the staggered pricing of \ct{C83} or the price-adjustment cost of \ct{R82}. Both pricing assumptions lead to the same linearized Phillips curve around the zero-inflation steady state, and therefore to the same simulations \cp{R95}. But the \name{C83} interpretation of $\x$ is easier to map to the data, so we use it for calibration. The parameter $\x$ indicates the share of firms that cannot update their price each period; it can be calibrated from microevidence on the frequency of price adjustments. If a share $\x$ of firms keep their price fixed each period, the average duration of a price spell is $1/(1-\x)$ \cp[p.~43]{G08}. In the microdata underlying the US Consumer Price Index, the mean duration of price spells is about 3 quarters \cp[Table~1]{NS13}. Hence, we set $1/(1-\x)=3$, which implies $\x=0.67$.

\subsection{Effects of Monetary Policy in the Short Run}

Price rigidity is a central concept in macroeconomic theory because it is a source of monetary nonneutrality. Here we explore how our pricing theory produces monetary nonneutrality. At this stage, we focus on the short-run effects of monetary policy, tracing how an unexpected and transitory shock to monetary policy permeates through the economy.

\paragraph{Analytical Results}  The dynamics of the textbook New Keynesian model around the steady state are governed by an IS equation, describing households' consumption-savings decisions, and a short-run Phillips curve, describing firms' pricing decisions. In the model with fairness, the same IS equation remains valid, but the Phillips curve is modified---because firms price differently.\footnote{Introducing fairness concerns into the New Keynesian model improves the realism of the Phillips curve but does not modify the IS equation. Yet the IS equation is also problematic: it notably creates several anomalies at the zero lower bound. Other behavioral elements have been introduced into the New Keynesian model to improve the realism of the IS equation. For instance, \ct{G16} assumes that households are inattentive to unusual events. Alternatively, \ct{MS18} assume that households derive utility from social status, which is measured by relative wealth.} 

The main difference is that the Phillips curve involves not only employment and inflation but also the perceived price markup, which itself obeys the following law of motion:

\begin{lemma}\label{l:monetary} In the New Keynesian model with fairness, the perceived price markup evolves according to
\begin{equation}
\wh{m^p}(t) = \g  \bs{\wh{\pi}(t) + \wh{m^p}(t-1)}.
\label{e:mphat}\end{equation}
Hence, the perceived price markup is a discounted sum of lagged inflation terms:
\begin{equation*}
\wh{m^p}(t) =\sum_{s=0}^{\infty}\g^{s+1} \wh{\pi}(t-s).
\end{equation*}\end{lemma}

The proof appears in Appendix~B.4; it is obtained by reworking the inference rule~\eqref{e:cpj}.

Equation \eqref{e:mphat} shows that the perceived price markup today tends to be high if inflation is high or if the past perceived markup was high. Past beliefs matter because people use them as a basis for their current beliefs. Inflation matters because people do not fully appreciate the effect of inflation on nominal marginal costs. Because of its autoregressive structure, the perceived price markup is fully determined by past inflation. 

As a result, the short-run Phillips curve involves not only forward-looking elements---expected future inflation and employment---but also backward-looking elements---past inflation.

\begin{proposition}\label{p:monetary} In the New Keynesian model with fairness, the short-run Phillips curve is
\begin{equation}
(1-\d\g) \wh{m^p}(t) - \l_1 \wh{n}(t)  = \d\g \E[t]{\wh{\pi}(t+1)} - \l_2 \E[t]{\wh{n}(t+1)},
\label{e:phillips}\end{equation}
where
\begin{align*}
\l_1 &\equiv (1+\eta) \frac{\e+(\e-1) \g  \ol{\f}}{\g  \ol{\f} \ol{\s}}\bs{1+\frac{(1-\d) \g}{1-\d\g}  \ol{\f}}\\
\l_2 &\equiv (1+\eta)  \d  \frac{\e+(\e-1) \ol{\f}}{\ol{\f}  \ol{\s}} \bs{1+\frac{(1-\d) \g}{1-\d\g}  \ol{\f}}.
\end{align*}
Accordingly, the short-run Phillips curve is hybrid, including both past and future inflation rates:
 \begin{equation*}
(1-\d\g) \sum_{s=0}^{\infty}\g^{s+1} \wh{\pi}(t-s) - \l_1 \wh{n}(t)  = \d\g \E[t]{\wh{\pi}(t+1)} - \l_2 \E[t]{\wh{n}(t+1)}.
\end{equation*}\end{proposition}

The proof appears in Appendix~B.4. It is obtained by log-linearizing firms' pricing equation~\eqref{e:em} around the steady state, and combining it with \eqref{e:n}---to link the price markup to employment---and with \eqref{e:e2} and \eqref{e:mphat}---to link the price elasticity of demand to inflation.

\paragraph{Simulation Results}  Next we simulate the dynamical response of our calibrated model to an unexpected and transitory monetary shock. Following the literature, we simulate dynamics around the zero-inflation steady state. 

We assume that the exogenous component $i_0(t)$ of the monetary-policy rule \eqref{e:taylor} follows an AR(1) process such that
\begin{equation*}
\wh{i_0}(t)=\m^i \cdot \wh{i_0}(t-1) - \z^i(t),
\end{equation*}
where the disturbance $\z^i(t)$ follows a white-noise process with mean zero, and $\m^i\in(0,1)$ governs the persistence of shocks. We set $\m^i = 0.75$, which corresponds to moderate persistence (\inp[p.~52]{G08}; \inp[p.~26]{G10}). We simulate the response to an initial disturbance of $\z^i(0) = 0.25\%$, which is an expansionary monetary shock. Without any inflation response, this shock would reduce the annualized interest rate by 1 percentage point.

\begin{figure}[p]
\includegraphics[scale=0.2,page=1]{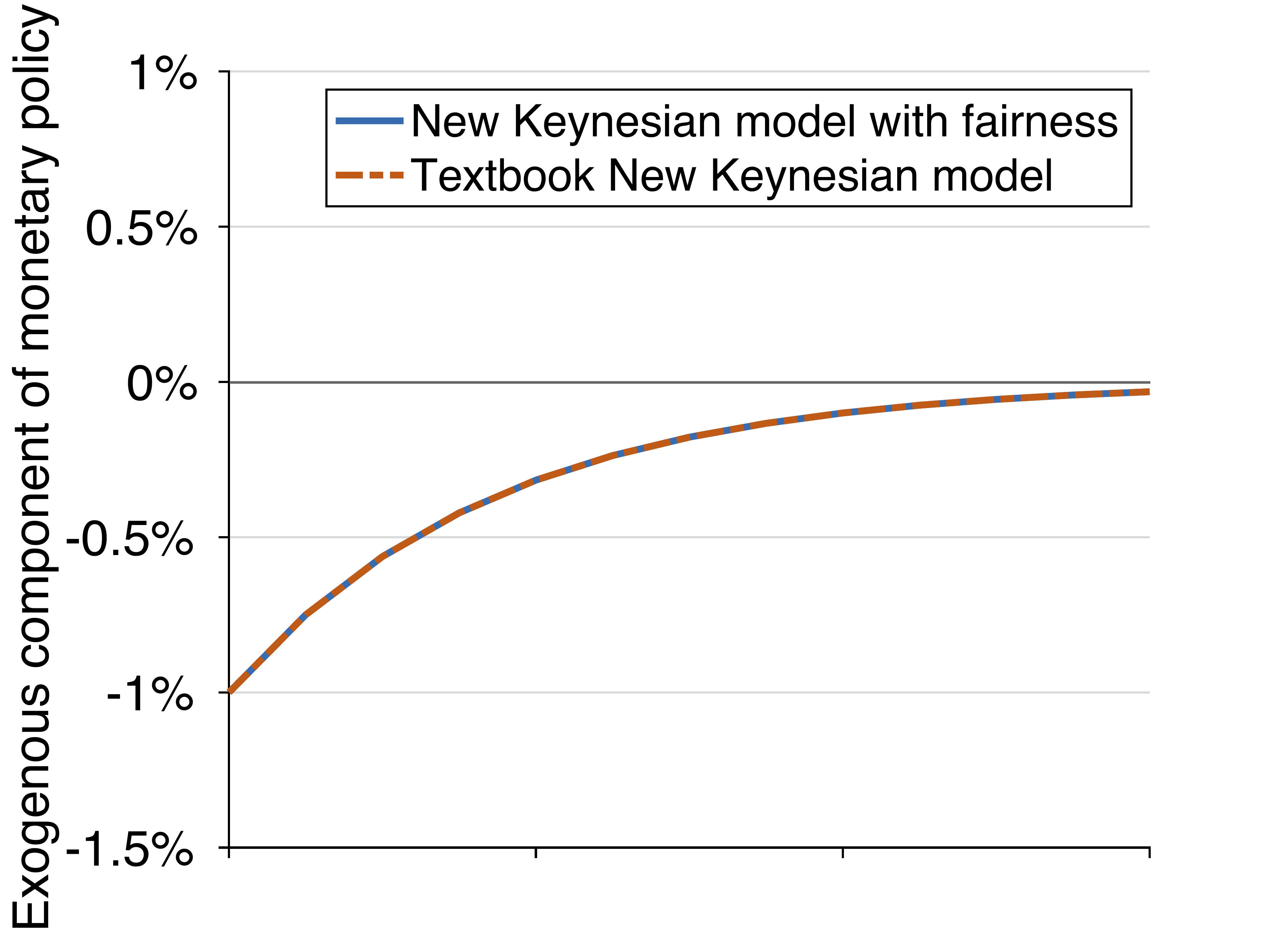}\hfill
\includegraphics[scale=0.2,page=2]{\pdf}\\
\includegraphics[scale=0.2,page=3]{\pdf}\hfill
\includegraphics[scale=0.2,page=4]{\pdf}\\
\includegraphics[scale=0.2,page=5]{\pdf}\hfill
\includegraphics[scale=0.2,page=6]{\pdf}\\
\caption{Effects of an expansionary monetary shock}
\note{This figure describes the response of the New Keynesian model with fairness (solid lines) to a decrease in the exogenous component of the monetary-policy rule \eqref{e:taylor} by 1 percentage point (annualized) at time 0. The exogenous component of monetary policy and inflation rate are deviations from steady state, measured in percentage points and annualized. The other variables are percentage deviations from steady state. For comparison, the figure also displays the response of the textbook New Keynesian model (dashed lines). The log-linearized equilibrium conditions used in the simulation of the model with fairness are presented in Appendix~B.4; those used in the simulation of the textbook model are in Appendix~C. The calibration of the two models is described in Table~\ref{t:calibration}.}
\label{f:monetary}\end{figure}

Figure~\ref{f:monetary} depicts the dynamical response to the expansionary monetary shock. The exogenous component of monetary policy and inflation rate are expressed as deviations from steady-state values, measured in percentage points and annualized (by multiplying by four the variables $\wh{i_0}(t)$ and $\wh{\pi}(t)$); all other variables are expressed as percentage deviations from steady-state values.

Loosening monetary policy raises inflation. Observing higher prices, customers underinfer the underlying increase in nominal marginal costs and thus perceive higher price markups. Firms respond to such perceptions by cutting their actual markups. The price markup falls by $1.4\%$, which raises output and employment by $0.7\%$. (Output and employment respond identically because the production function is calibrated to be linear.)

\paragraph{Comparison with Microevidence}  In our model, when consumers observe inflation, they misperceive price markups as higher and transactions as less fair, which lowers their consumption utility and triggers a feeling of displeasure. The survey responses collected by \ct{S96} agree with these predictions. Among 120 respondents in the United States, 85\% report that they dislike inflation because when they ``go to the store and see that prices are higher,'' they ``feel a little angry at someone'' (p.~21). The most common perceived culprits are ``manufacturers,'' ``store owners,'' and ``businesses,'' whose transgressions include ``greed'' and ``corporate profits'' (p.~25). Thus, in our model as in the real world, consumers perceive higher markups during inflationary periods and are angered by them. 

Conversely, upon observing deflation, consumers in the model would believe that price markups are lower and transactions more fair, which would boost their consumption utility and trigger a feeling of happiness. Hence, our model predicts not only that consumers dislike inflation, but also that they enjoy deflation. An opinion poll conducted by the Bank of Japan between 2001 and 2017 paints this pattern (Table~\ref{t:japan}). During this period, Japan alternated between inflation and deflation. Of the 68,000 respondents facing price increases, only 3\% see that as a favorable development, whereas 84\% see it as unfavorable. In contrast, of the 18,000 respondents facing price decreases, 43\% regard that as a favorable development, and only 22\% regard it as unfavorable.

\begin{table}[t]
\caption{Opinions about price movements in Japan, 2001--2017}
\footnotesize\begin{tabular*}{\textwidth}{@{\extracolsep{\fill}}lcccc}\toprule 
& &\multicolumn{3}{c}{Opinion about perceived price change}  \\ 
\cmidrule{3-5}
Perceived price change	& Respondents & Favorable & Neutral & Unfavorable \\ 
\midrule 
Prices have gone up 		& 68,491		& $2.5\%$ 	& $13.0\%$ & $83.7\%$ \\ 
Prices have gone down  	&  18,257	& $43.0\%$	& $34.2\%$ & $21.9\%$ \\ 
\bottomrule\end{tabular*}
\note{Data come from the 60 waves of the Opinion Survey on the General Public's Mindset and Behavior conducted by the Bank of Japan between September 2001 and December 2017. (Earlier data is not online and has therefore been excluded.) The survey was conducted quarterly on a random sample of 4,000 adults living in Japan, with a $57.2\%$ average response rate. Respondents answered the following question: ``How do you think prices (defined as overall prices of goods and services you purchase) have changed compared with one year ago?'' (question 10, 11, 12, or 13, depending on the survey). Respondents who answered ``prices have gone up significantly'' or ``prices have gone up slightly'' are described on the first row of the table. Respondents who answered ``prices have gone down significantly'' or  ``prices have gone down slightly'' are described on the second row of the table.  The remainder, who answered ``prices have remained almost unchanged,'' do not feature in the table. Those who answered that prices had gone up then answered ``How would you describe your opinion of the price rise?'' (question 10, 11, 12, or 13, depending on the survey, and only after June 2004). The third column gives the share of those respondents who answered ``rather favorable,'' the fourth column the share who answered ``neither favorable nor unfavorable,'' and the fifth column the share who answered ``rather unfavorable.''  Those who answered that prices had gone down then answered ``How would you describe your opinion of the price decline?'' (question 10, 11, 12, 13, or 15, depending on the survey). The third, fourth, and fifth column give the share of those respondents who answered ``rather favorable,'' ``neither favorable nor unfavorable,'' and ``rather unfavorable.'' Detailed survey results are available at \url{http://www.boj.or.jp/en/research/o_survey/index.htm/}.}
\label{t:japan}\end{table}

\paragraph{Comparison with Macroevidence}  Monetary policy is nonneutral in the model because monetary shocks influence output and employment. The nonneutrality of monetary policy is well documented; the evidence is summarized by \ct{CEE99} and \ct[Section~3]{R15}. Furthermore, the effect of monetary policy is mediated by a hybrid Phillips curve, which is realistic as both past inflation and expected future inflation enter significantly in estimated New Keynesian Phillips curve \cp[Table~2]{MPS14}.

In fact the response of output to a monetary shock is broadly the same in the model as in US data. First, the shape of the response is similar, as output is estimated to respond to monetary shocks in a hump-shaped fashion \cp[Figures~1--4]{R15}. Second, the amplitude of the response is comparable. After a one-percentage-point decrease of the nominal interest rate, the literature estimates that output increases between $0.6\%$ and $5\%$, with a median value of $1.6\%$ \cp[Table~1]{R15}; and using a range of methods and samples, \ct[Table~2]{R15} estimates that output increases between $0.2\%$ and $2.2\%$, with a median value of $0.8\%$. In our simulation, output rises by $0.7\%$ when the exogenous component of monetary policy decreases by~1 percentage point---close to Ramey's median estimate.

After a monetary shock, price markup and output move in opposite directions; the same would be true after other aggregate-demand shocks. Moreover, aggregate-demand shocks explain most business-cycle fluctuations \cp{G99,BFK06}. Accordingly, our model predicts that price markups are countercyclical. And indeed, price markups seem countercyclical in the data \cp{RW99,BKM18}.

The main discrepancy between our model and US evidence concerns inflation. Whereas inflation in the United States responds in a delayed and gradual way to monetary shocks \cp[Figures~1--4]{R15}, both our model and the textbook model predict an immediate response.

\paragraph{Comparison with the Textbook New Keynesian Model}  In both our model and the textbook model, looser monetary policy leads to higher inflation and lower markups, boosting employment and output. Beyond these similarities, the two models differ on several counts.

First, the textbook model's short-run Phillips curve is purely forward-looking, so it does not include the backward-looking elements found in US data and present in the fairness model. Of course, other variations of the textbook model append backward-looking components to the Phillips curve; for example, having firms index their prices to past inflation in periods when they cannot reset their prices \cp{CEE05}.

Second, the textbook model cannot produce the positive correlation between perceived price markup and inflation that occurs in the fairness model, and that rationalizes the survey findings by \ct{S96} and the Bank of Japan. This is because households in the textbook model correctly infer that price markups are lower when they see higher inflation.

Third, the textbook model cannot produce the hump-shaped response of output observed in US data and predicted by the fairness model, since it does not include any backward-looking element. The fairness model, by contrast, includes a backward-looking element in the form of the perceived price markup $\wh{m^p}(t)$, which enters the Phillips curve \eqref{e:phillips} and depends on the past via \eqref{e:mphat}. It is well understood that hump-shaped impulse responses can be obtained by inserting backward-looking elements---for instance, by assuming that consumers form habits \cp{F00,CEE05}. Under that assumption, consumers' behavior depends on their past consumption, which then enters the IS curve and generates hump-shaped responses.

Fourth, the response of output in the textbook model is about one third the size of that in the fairness model, and much smaller than in US data. Despite both models being calibrated through microevidence on price dynamics, monetary shocks are more amplified in the fairness model.

\subsection{Effects of Technology Shocks}

\begin{figure}[p]
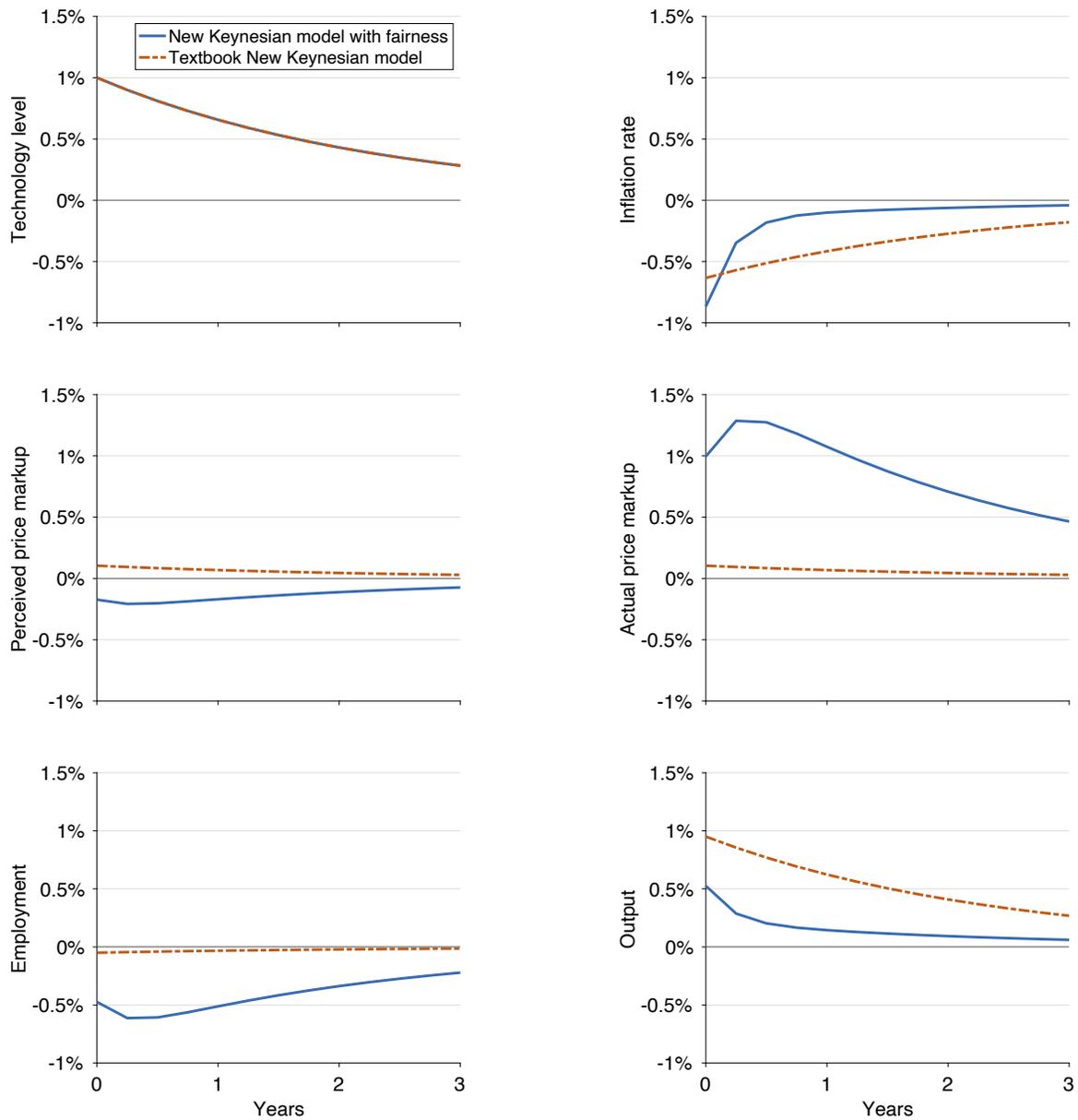

\includegraphics[scale=0.2,page=7]{\pdf}\hfill
\includegraphics[scale=0.2,page=8]{\pdf}\\
\includegraphics[scale=0.2,page=9]{\pdf}\hfill
\includegraphics[scale=0.2,page=10]{\pdf}\\
\includegraphics[scale=0.2,page=11]{\pdf}\hfill
\includegraphics[scale=0.2,page=12]{\pdf}\\
\caption{Effects of a positive technology shock}
\note{This figure describes the response of the New Keynesian model with fairness (solid lines) to a 1\% increase in technology at time $0$. The inflation rate is a deviation from steady state, measured in percentage points and annualized. The other variables are percentage deviations from steady state. For comparison, the figure also displays the response of the textbook New Keynesian model (dashed lines). The log-linearized equilibrium conditions used in the simulation of the model with fairness are presented in Appendix~B.4; those used in the simulation of the textbook model are in Appendix~C. The calibration of the two models is described in Table~\ref{t:calibration}.}
\label{f:technology}\end{figure}

The price rigidity arising from fairness concerns allows monetary policy to influence real variables such as employment and output. It also affects the transmission of nonpolicy shocks to the economy. Here we illustrate the effects of a technology shock---the most widely studied nonpolicy shock in modern macroeconomics---on the economy. 

\paragraph{Simulation Results}  We simulate the dynamical response of our calibrated model to an unexpected and transitory technology shock, once again around the zero-inflation steady state. We assume that the logarithm of technology $A(t)$ in the production function \eqref{e:yj} follows an AR(1) process such that
\begin{equation*}
\wh{a}(t)=\m^a \cdot \wh{a}(t-1)+ \z^a(t),
\end{equation*}
where the disturbance $\z^a(t)$ follows a white-noise process with mean zero, and $\m^a\in(0,1)$ governs the persistence of shocks. We set $\m^a = 0.9$, which is typical \cp[p.~55]{G08}. We simulate the response to an initial disturbance of $\z^a(0) = 1\%$.

Figure~\ref{f:technology} displays the response to the positive technology shock. The inflation rate is expressed as a deviation from its steady-state value, measured in percentage points and annualized (by multiplying by four the variable $\wh{\pi}(t)$); all other variables are expressed as percentage deviations from their steady-state values.

The increase in technology reduces marginal costs, pulling down inflation. Observing lower prices, customers underinfer the underlying decrease in marginal costs and thus perceive lower price markups and fairer transactions. The improvement in perceived fairness decreases the price elasticity of the demand for goods. Firms best respond by raising their markups. The price markup increases by $1.3\%$ at the peak, which depresses employment by $0.7\%$. Despite the drop in employment, output initially increases by $0.5\%$ due to improved technology.

\paragraph{Comparison with Macroevidence}  In our model, an increase in technology leads to higher output but lower employment. This prediction conforms to much of the evidence from US data \cp{GR04,BFK06,FR09}. Our model also predicts that inflation falls after the increase in technology, as documented by \ct[Figure~4]{BFK06}. Finally, in the model, price markups and output are positively correlated under technology shocks. \ct{NR13} report evidence consistent with this prediction.

\paragraph{Comparison with the Textbook New Keynesian Model}  The similarities and differences between the two models identified under monetary shocks also apply under technology shocks. The two models predict the same direction of change in inflation, price markup, employment, and output. There are three main differences. First, the fairness model produces a hump-shaped response of employment to technology shocks, which the textbook model does not. Second, the fairness model produces a negative correlation between perceived and actual price markups, whereas the textbook model does not distinguish between the two. Last, in response to a positive technology shock,  employment falls more in the fairness model than in the textbook model; consequently, output increases less in the fairness model than in the textbook model.

\subsection{Effects of Monetary Policy in the Long Run}

Our pricing theory implies that monetary policy is nonneutral in the short run, so a transitory monetary shock affects employment. Here we develop another implication of the theory: monetary policy is nonneutral in the long run, so a change in the steady-state inflation rate affects steady-state employment. Thus, the theory generates a nonvertical long-run Phillips curve.

We study the long-run effects of monetary policy by comparing the steady-state equilibria induced by different values of the exogenous component $\ol{i_0}$ in the monetary-policy rule \eqref{e:taylor}. In steady state the real interest rate equals the discount rate  $\r \equiv -\ln(\d)$; therefore, by choosing $\ol{i_0}$, monetary policy perfectly controls steady-state inflation:
\begin{equation*}	
\ol{\pi}  = \frac{\r-\ol{i_{0}}}{\p-1}.
\end{equation*}
To obtain zero inflation, it suffices to set $\ol{i_0}=\r$; to obtain higher inflation, it suffices to reduce $\ol{i_0}$.

\paragraph{Acclimation}  \ct[p.~730]{KKT86} hypothesize that ``any stable state of affairs tends to become accepted eventually''. We adapt their idea by assuming that people partially acclimate to the steady-state inflation rate, generalizing the fairness function \eqref{e:f} to 
\begin{equation}
F(M^p) = 1 - \t \cdot(M^p - M^f),
\label{e:facclim}\end{equation} 
where $M^f$ is the fair markup resulting from acclimation. We assume that the fair markup is the weighted average of the standard markup $\e/(\e-1)$ and the steady-state perceived markup $\ol{M^p}$:
\begin{equation}
M^f = \c \cdot \ol{M^p} + (1-\c) \cdot \frac{\e}{\e-1}.
\label{e:mf}\end{equation}
The parameter $\c \in [0,1]$ measures acclimation: when $\c=0$, there is no acclimation; when $\c=1$, there is full acclimation, so people do not mind whatever is happening in steady state; when $\c\in(0,1)$, people may be permanently satisfied or dissatisfied in steady state, but less than when $\c=0$.\footnote{This specification does not change anything at the zero-inflation steady state. With zero inflation, $\ol{M^p} = \e/(\e-1)$, so $M^f = \e/(\e-1)$ for any $\c$. Therefore, the fairness function~\eqref{e:facclim} simplifies to the function~\eqref{e:f} for any~$\c$.}

\paragraph{Analytical Results}  In steady state, the rate of inflation determines the perceived price markup, fairness factor, and elasticity of the fairness function:

\begin{lemma}\label{l:longrun} In the New Keynesian model with fairness, the steady-state perceived price markup is a strictly increasing function of steady-state inflation:
\begin{equation*}
\ol{M^{p}}(\ol{\pi}) = \frac{\e}{\e-1} \cdot \exp{\frac{\g}{1-\g}  \ol{\pi}}.
\end{equation*}
Hence, the steady-state fairness factor is a weakly decreasing function of steady-state inflation:
\begin{equation*}
\ol{F}(\ol{\pi}) = 1-\t \cdot (1-\c)\cdot\bs{\ol{M^p}(\ol{\pi})-\frac{\e}{\e-1}}.	
\end{equation*}
Accordingly, the steady-state elasticity of the fairness function is a strictly increasing function of steady-state inflation:
\begin{equation*}
\ol{\f}(\ol{\pi}) = \frac{\t \cdot \ol{M^p}(\ol{\pi})}{\ol{F}(\ol{\pi})}.	
\end{equation*}\end{lemma}

The proof involves manipulating the inference mechanism~\eqref{e:cpj} to obtain $\ol{M^p}$, and using \eqref{e:facclim} and \eqref{e:mf} to obtain $\ol{F}$ and $\ol{\f}$. It appears in Appendix~B.3.

The lemma shows that in steady state households perceive higher price markups when inflation is higher. Households understand that in steady state nominal marginal costs grow at the inflation rate, but because of subproportional inference, they misjudge the level of those costs and thus of price markups. Since perceived price markups are higher when inflation is higher, the fairness factor is lower---except when consumers are fully acclimated ($\c=1$), in which case the fairness factor is always one. Last, the elasticity of the fairness function is higher when inflation is higher.

From Lemma~\ref{l:longrun}, we infer that the long-run Phillips curve is upward sloping:

\begin{proposition}\label{p:longrun} In the New Keynesian model with fairness, the steady-state price markup is a strictly decreasing function of steady-state inflation:
\begin{equation}
\ol{M}(\ol{\pi})=1+\frac{1}{\e-1}\cdot\frac{1}{1+\frac{(1-\d)\g}{1-\d\g}\ol{\f}(\ol{\pi})}.
\label{e:mss}\end{equation}
Hence, steady-state employment is a strictly increasing function of steady-state inflation:
\begin{equation*}
\ol{N}=\bs{\frac{(\n-1) \a}{\n}\cdot\frac{1}{\ol{M}(\ol{\pi})}}^{1/(1+\eta)}.
\end{equation*}
Thus, the long-run Phillips curve is not vertical (fixed $\ol{N}$) but upward sloping.
\end{proposition}

The proof appears in Appendix~B.3; its main step is reworking \eqref{e:em} in steady state to obtain $\ol{M}$.

Because the long-run Phillips curve slopes upward for any degree of acclimation, monetary policy is nonneutral in the long run. In fact, \eqref{e:mss} has the same structure as \eqref{e:m}, so the New Keynesian model operates like the monopoly model. After an increase in inflation, households underappreciate the increase in nominal marginal costs, so they partly attribute the higher prices to higher markups, which they find unfair. Since perceived markups are higher, the price elasticity of demand increases, leading firms to reduce their actual markups.

Last, we obtain comparative statics on the slope of the long-run Phillips curve:

\begin{corollary}\label{c:longrun} In the New Keynesian model with fairness, around the zero-inflation steady state, the slope of the long-run Phillips curve is
\begin{equation*}
\od{\ol{\pi}}{\ln(\ol{N})} = \frac{1+\eta}{1-\d}\cdot \frac{(1-\g)(1-\d\g)}{\g^2}\cdot \frac{\e-1}{\t} \cdot \frac{\bs{1+\frac{(1-\d)\g}{1-\d\g}\t}\bs{\bp{1+\frac{(1-\d)\g}{1-\d\g}\t}\e-1}}{\bs{1+(1-\c)\t}\e-1}.
\end{equation*}
The slope increases with the competitiveness of the goods market ($\e$) and degree of acclimation ($\c$); it decreases with the concern for fairness ($\t$) and degree of underinference ($\g$).\end{corollary}

The proof is relegated to Appendix~B.3; it builds on the results in Lemma~\ref{l:longrun} and Proposition~\ref{p:longrun}.

The impact of competitiveness, fairness concern, and degree of underinference on the slope of the long-run Phillips curve is reminiscent of the impact of these parameters on the cost passthrough in the monopoly model (see Corollary~\ref{c:subproportional}). The impact of the degree of acclimation is easily understandable. With more acclimation, perceived fairness ($\ol{F}$) depends less on inflation, because consumers adapt more to different inflation rates. As a result, the elasticity of the fairness function ($\ol{\f}$) depends less on inflation, and so the Phillips curve \eqref{e:mss} is steeper.  

The interpretation of the corollary is that lower competitiveness on the goods market, lower acclimation, stronger concern for fairness, and stronger underinference flatten the long-run Phillips curve---thus strengthening the long-run effects of monetary policy.

\begin{figure}[p]
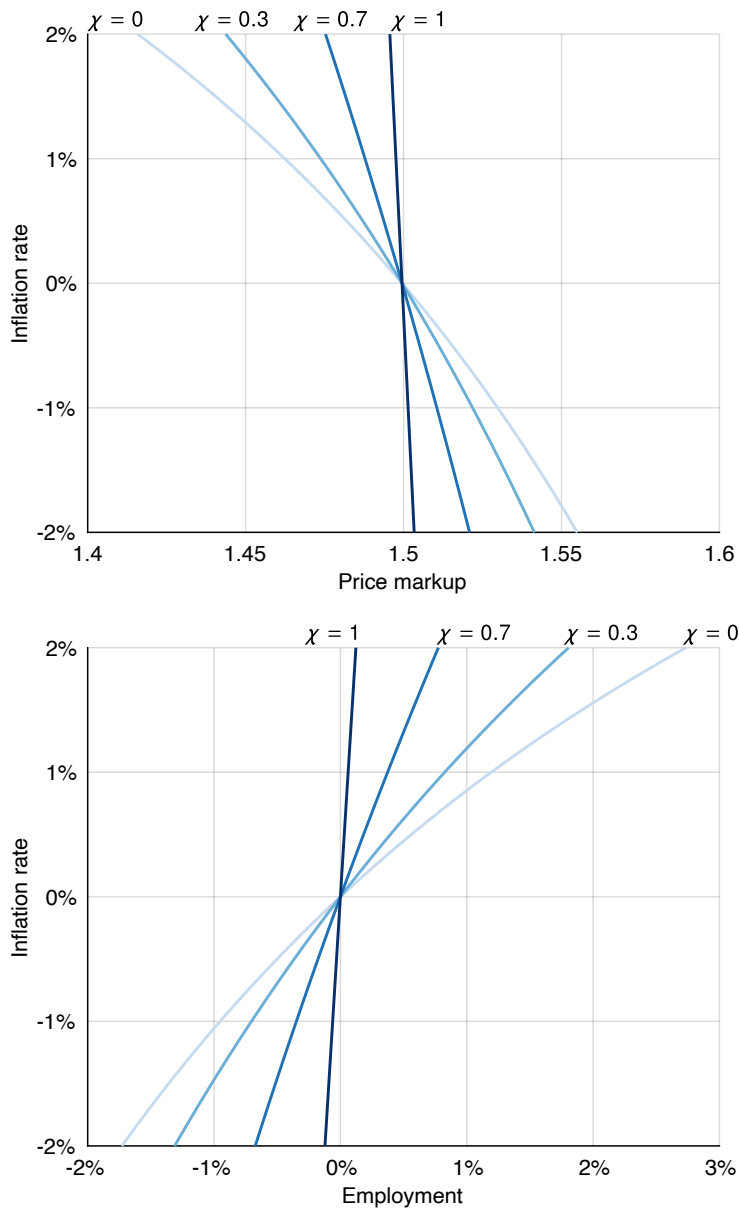

\includegraphics[scale=0.3,page=13]{\pdf}
\includegraphics[scale=0.3,page=14]{\pdf}
\caption{Long-run Phillips curves for various degrees of acclimation}
\note{The top panel gives the relationship between steady-state inflation rate and steady-state price markup. The bottom panel gives the relationship between steady-state inflation rate and steady-state employment. In both panels, inflation is measured as an annual rate. In the bottom panel, employment is measured as a percentage deviation from employment in the zero-inflation steady state. These long-run Phillips curve are constructed using the expressions in Proposition~\ref{p:longrun} under the calibration in Table~\ref{t:calibration}, for various degrees of acclimation: $\c=0$ (no acclimation), $\c = 0.3$, $\c = 0.7$, and $\c=1$ (full acclimation).}
\label{f:longrun}\end{figure}

\paragraph{Simulation Results}  To quantify long-run monetary nonneutrality, we compute the long-run Phillips curve in our calibrated model. Figure~\ref{f:longrun} displays two versions of the curve: one describes the relationship between steady-state inflation rate and steady-state price markup, and the other the relationship between steady-state inflation rate and steady-state employment. Absent microevidence on acclimation, we compute the Phillips curve for various degrees of acclimation. 

With full acclimation ($\c=1$), the Phillips curve is almost vertical, so in the long run inflation barely affects price markup and employment. For instance, increasing the inflation rate from 0\% to 1\% only raises employment by $0.06\%$. With partial acclimation, the Phillips curve becomes flatter. With an acclimation of $\c=0.7$, the same increase in inflation raises employment by $0.4\%$; and with a lower acclimation of $\c=0.3$, it raises employment by $0.8\%$. Finally, with no acclimation ($\c=0$), the Phillips curve is even flatter, and inflation has a larger effect on price markup and employment. Then, increasing inflation from 0\% to 1\% raises employment by $1.2\%$.

\paragraph{Comparison with Macroevidence}  The property that higher steady-state inflation leads to higher steady-state employment is consistent with evidence that higher average inflation leads to lower average unemployment. \ct[Table~1]{KW94} find in US data that a permanent increase in inflation by 1 percentage point reduces the unemployment rate between $0.2$ and $1.3$ percentage points, depending on the period and identification strategy. \ct{KW97} confirm these findings, while highlighting the uncertainty surrounding the Phillips curve's slope.

Quantitatively, the findings by \ct{KW94} also agree with our model's predictions. Abstracting from possible changes in labor force participation, their findings imply that increasing inflation by 1 percentage point raises employment by $0.2\%$ to $1.3\%$. This magnitude matches the simulation results for a degree of acclimation between $\c=0$ and $\c=0.7$.

The mechanism behind the upward-sloping long-run Phillips curve is that higher steady-state inflation lowers steady-state price markups. There is evidence that this mechanism operates. \ct{B92} uncovers that in the US retail sector, higher average inflation leads to lower average markup. \ct{BR05} reach the same conclusion using aggregate US data.

\paragraph{Comparison with the Literature} In the textbook New Keynesian model, steady-state inflation also affects the price markup and employment. With \name{R82} pricing, higher steady-state inflation leads to a lower price markup and higher output, as in our model \cp[Figure~1]{AR12}. With \name{C83} pricing, the opposite occurs: higher steady-state inflation leads to a higher price markup and lower output, which appears counterfactual \cp[Figure~2]{AR12}.\footnote{Although the New Keynesian models with \name{R82} pricing and \name{C83} pricing coincide around the zero-inflation steady state, they differ elsewhere.}

Our mechanism for an upward-sloping long-run Phillips curve complements the mechanism proposed by \ct{ADP96} and \ct{BR11}: that higher inflation reduces the likelihood that firms facing negative shocks be forced by downward nominal wage rigidity to fire workers. The two mechanisms may have the same psychological origin as downward nominal wage rigidity seems to stem from workers' fairness concerns \cp{B07}.

\section{Conclusion}\label{s:conclusion}

This paper develops a theory of pricing to fairness-minded customers that revolves around two assumptions. First, customers derive more utility from a good priced at a low markup---perceived as fairly priced---than one priced at a high markup---perceived as unfairly priced. Second, customers infer firms' hidden marginal costs from firms' prices in a subproportional manner: they infer too little, and to the extent that they do infer, they misperceive marginal costs as proportional to prices. These assumptions conform to copious evidence collected from customers and firms. 

The main implication of the theory is price rigidity: the passthrough of marginal costs into prices is strictly less than one. When the theory is embedded into a New Keynesian model, price rigidity leads to the nonneutrality of monetary policy, both in the short run and in the long run. Furthermore, we are able to calibrate our two psychological parameters---concern for fairness and degree of underinference---from microevidence, just as any other parameter of the model. When simulating the calibrated model, we obtain realistic impulse responses of output and employment to monetary shocks: the responses are hump-shaped and have the appropriate amplitude. We also obtain realistic impulse responses to technology shocks: a transitory improvement in technology leads to higher output but lower employment.

The paper delineates a mechanism through which fairness affects a market economy. Hidden information and underinference play crucial roles. When costs are observable, or when costs are hidden but customers infer them rationally from prices, our model with fairness is isomorphic to a model without fairness. Only when costs are hidden and customers infer subproportionally does fairness affect the qualitative properties of equilibrium, such as by creating price rigidity. Another key ingredient to our theory is that fairness modifies the price elasticity of demand, which allows fairness to sway large markets---a feature not shared by common approaches to fairness \cp{DHK11,S07}.

Our model helps bridge a gap between the public's attitude toward inflation and the harm from inflation described by macroeconomic models. \ct[p.~519]{R02} argues that ``There is a wide gap between the popular view of inflation and the costs of inflation that economist can identify. Inflation is intensely disliked. In periods when inflation is moderately high in the United States, for example, it is often cited in opinion polls as the most important problem facing the country. It appears to have an important effect on the outcome of Presidential elections. Yet, economists have difficulty in identifying substantial costs of inflation.''  Our model contributes to explaining such intense dislike for inflation. 

Finally, we hope that our theory might be fruitfully applied to the study of optimal monetary policy. Since its microfoundations match the motivations of real-world customers and firms, as well as their real-world reactions to inflation and deflation, the theory should underpin a more accurate welfare function that could enhance the design of monetary policy.

\bibliography{\bib}

\appendix

\section{Derivations for the Monopoly Model}\label{a:monopoly}

We derive several of the monopoly results stated in Section~\ref{s:monopoly}. In particular, we provide proofs for Lemma~\ref{l:phi}, Proposition~\ref{p:subproportional}, and Corollary~\ref{c:subproportional}.

\subsection{Properties of the Profit Function}\label{a:foc}

We show that in all the cases treated in Section~\ref{s:monopoly}, the first-order condition gives the maximum of the monopoly's profit function. This is because the profit function is unimodal.

The monopoly chooses a price $P>C$ to maximize profits 
\begin{equation*}
V(P) = (P-C) \cdot Y^{d}(P). 
\end{equation*}
The profit function is differentiable. Its derivative is
\begin{equation*}
V'(P) = Y^{d}+\bp{P-C} \od{Y^{d}}{P} =Y^{d}- \bp{P-C} \frac{Y^d}{P} E(P),
\end{equation*}
where 
\begin{equation*}
E(P) \equiv -\odl{Y^{d}}{P} = -\frac{P}{Y^d}\cdot\od{Y^{d}}{P}
\end{equation*}
is the price elasticity of demand. Hence the derivative of the profit function satisfies
\begin{equation}
V'(P) =Y^{d}(P) \bs{1 - \frac{P-C}{P} E(P)}.
\label{e:dvdp}\end{equation}
We now study the properties of the derivative \eqref{e:dvdp} in the various cases considered in Section~\ref{s:monopoly}.

\paragraph{No Fairness Concerns} Without fairness concerns, the price elasticity of demand is $E=\e$ (Section~\ref{s:nofairness}). Hence the derivative \eqref{e:dvdp} becomes
\begin{equation*}
V'(P) =Y^{d}(P) \bs{1 - \e\frac{P-C}{P}}.
\end{equation*}
The function $P\mapsto (P-C)/P$ is strictly increasing from 0 to 1 as $P$ increases from $C$ to $+\infty$, so the term in square brackets is strictly decreasing from 1 to $1-\e<0$ as $P$ increases from $C$ to $+\infty$. Hence, the term in square brackets has a unique root $P^*$ on $(C,+\infty)$, is positive for $P<P^*$, and is negative for $P>P^*$. Since $Y^d(P)>0$, these properties transfer to the derivative of the profit function: $V'(P)>0$ for $P\in(C,P^*)$, $V'(P)=0$ at $P=P^*$, and $V'(P)<0$ for $P\in(P^*,+\infty)$. We conclude that the profit function is unimodal, and its maximum $P^*$ is the unique solution to the first-order condition $V'(P)=0$.

\paragraph{Fairness Concerns and Observable Costs} With fairness concerns and observable costs, the price elasticity of demand is $E=\e+(\e-1)\f(P/C)$ (Section~\ref{s:observable}). The profit function is now defined for $P\in (C, M^h \cdot C)$. The derivative \eqref{e:dvdp} becomes
\begin{equation*}
V'(P) =Y^{d}(P) \bs{1 - \frac{P-C}{P} \cdot \bc{\e+(\e-1)\f(P/C)}}.
\end{equation*}
Again, the function $P\mapsto (P-C)/P$ is strictly increasing from 0 to 1 as $P$ increases from $C$ to $+\infty$. The elasticity of the fairness function $\f(P/C)$ is strictly increasing from $\f(1)>0$ to $+\infty$ as $P$ increases from $C$ to $M^h \cdot C$ (Lemma~\ref{l:phi}). Hence the term in square brackets is strictly decreasing from 1 to $-\infty$ as $P$ increases from $C$ to $M^h \cdot C$. This implies that the term in square brackets has a unique root $P^*$ on $(C,M^h \cdot C)$, is positive for $P<P^*$, and is negative for $P>P^*$. Following the same argument as in the previous case, we conclude that the profit function is unimodal, and its maximum $P^*$ is the unique solution to the first-order condition $V'(P)=0$.

\paragraph{Fairness Concerns and Rational Inference of Costs} With fairness concerns and rational inference of marginal costs, the price elasticity of demand is again $E=\e$ (Section~\ref{s:rational}). Hence, as in the case of no fairness concerns, the profit function is unimodal and its maximum is the unique solution to the first-order condition $V'(P)=0$.

\paragraph{Fairness Concerns and Subproportional Inference of Costs} With fairness concerns and subproportional inference of costs, the price elasticity of demand is $E=\e+(\e-1)\g\f(M^p(P))$ (Section~\ref{s:subproportional}). The profit function is now defined for $P\in (C, P^b)$, where the upper bound is defined by 
\begin{equation}
P^b = \frac{\e}{\e-1} (M^h)^{1/\g} C^b.
\label{e:pb}\end{equation}
The price $P^b$ is such that at $P^b$, the perceived markup reaches the upper bound of the domain of the fairness function:  $M^p(P^b)=M^h$. We know that $P^b > C$ because $C^b > (\e-1) \cdot (M^h)^{-1/\g} \cdot C/\e$ (Definition~\ref{d:subproportional}). The derivative \eqref{e:dvdp} becomes
\begin{equation*}
V'(P) =Y^{d}(P) \bs{1 - \frac{P-C}{P} \cdot \bc{\e+(\e-1)\g\f(M^p(P))}}.
\end{equation*}
Again, the function $P\mapsto (P-C)/P$ is strictly increasing from 0 to 1 as $P$ increases from $C$ to $+\infty$. The perceived markup $M^p(P)$ is strictly increasing from $M^p(C)>0$ to $M^h$ as $P$ increases from $C$ to $P^b$ (Lemma~\ref{l:subproportional}). Hence, the elasticity of the fairness function $\f(M^p(P))$ is strictly increasing from $\f(M^p(C))>0$ to $+\infty$ as $P$ increases from $C$ to $P^b$ (Lemma~\ref{l:phi}). Since $\g>0$, we infer that the term in square brackets is strictly decreasing from 1 to $-\infty$ as $P$ increases from $C$ to $P^b$. Thus the term in square brackets has a unique root $P^*$ on $(C,P^b)$, is positive for $P<P^*$, and is negative for $P>P^*$. Following the same argument as in the previous cases, we conclude that the profit function is unimodal, and its maximum $P^*$ is the unique solution to the first-order condition $V'(P)=0$.

\subsection{Proof of Lemma~\ref{l:phi}}\label{a:lemma}

By definition, the elasticity of the fairness function is given by
\begin{equation*}
\f(M^p) = - M^p \cdot \frac{F'(M^p)}{F(M^p)}. 
\end{equation*}
The properties of the fairness function $F$ listed in Definition~\ref{d:fairness} indicate that $F(M^p)>0$ and $F'(M^p)<0$, so $\f(M^p)>0$. 

The properties also indicate that $F>0$ is decreasing in $M^p$, and that $F'<0$ is decreasing in $M^p$ (as $F$ is concave in $M^p$). Thus, both $1/F>0$ and $-F'>0$ are increasing in $M^p$, which implies that $\f$ is strictly increasing in $M^p$. 

Next, the properties indicate that $F(0)>0$ and $F'(0)$ is finite, so $\lim_{M^p\to 0} \f(M^p) = 0$. And they indicate that $F(M^h)=0$ while $M^h>0$ and $F'(M^h)<0$, so $\lim_{M^p\to M^h} \f(M^p) = +\infty$. 

Last, the superelasticity of the fairness function is given by
\begin{equation*}
\s = M^p \cdot \frac{\f'(M^p)}{\f(M^p)}.
\end{equation*}
Since $\f(M^p)>0$ and $\f'(M^p)>0$, it is clear that $\s>0$.

\subsection{Proof of Proposition \ref{p:subproportional}}\label{a:proposition}

\paragraph{Markup} Since customers care about fairness and infer subproportionally, the price elasticity of demand is $E = \e+(\e-1)  \g   \f(M^p(P))$. Moreover, the monopoly's optimal markup is 
\begin{equation*}
M = \frac{E}{E-1} = 1 + \frac{1}{E-1}. 
\end{equation*}
Combining these equations yields the markup
\begin{equation}
M =1+\frac{1}{\e-1}\cdot \frac{1}{1+\g \f(M^p(M \cdot C))}.
\label{e:ma}\end{equation}
In \eqref{e:ma} we have used the relationship between the price and markup: $P = M \cdot C$.

Toward showing that \eqref{e:ma} admits a unique solution, we introduce the price $P^b$ defined by \eqref{e:pb} and the markup $M^b = P^b/C >1$. Since $P = M \cdot C$, $P$ strictly increases from $0$ to $P^b$ when $M$ increases from $0$ to $M^b$. Next, Lemma~\ref{l:subproportional} shows that $M^p(P)$ strictly increases from 0 to $M^h$ when $P$ increases from $0$ to $P^b$. Last, Lemma~\ref{l:phi} indicates that $\f(M^p)$ strictly increases from $0$ to $\infty$ when $M^p$ increases from 0 to $M^h$. As $\g>0$, we conclude that when $M$ increases from $0$ to $M^b>1$, the right-hand side of \eqref{e:ma} strictly decreases from $\e/(\e-1)$ to $1$. Hence, \eqref{e:ma} has a unique solution $M\in\bs{0,M^b}$, implying that the markup exists and unique. Given the range of values taken by the right-hand side of \eqref{e:ma}, we also infer that $M \in (1, \e/(\e-1))$.

\paragraph{Passthrough} We now compute the cost passthrough, $\b= \odlx{P}{C}$. The equilibrium price is $P= M(M^p(P)) \cdot C$, where the markup $M(M^p)$ is given by \eqref{e:ma}. Using this price equation, we obtain
\begin{equation*}
 \b = \odl{M}{M^p}\cdot \odl{M^p}{P}\cdot\odl{P}{C}+1.
\end{equation*} 
Since $\odlx{M^p}{P}=\g$ (Lemma~\ref{l:subproportional}) and $\odlx{P}{C}=\b$ (by definition), we get
\begin{equation}
 \b = \frac{1}{1-\g \odl{M}{M^p}}.
\label{e:beta0}\end{equation} 

Our next step is to compute the elasticity of $M(M^p)$ with respect to $M^p$ from \eqref{e:ma}:
\begin{equation*}
-\odl{M}{M^p} = -\frac{1}{M}\cdot \od{M}{\ln(M^p)} = \frac{1}{M}\cdot \frac{1}{\e-1}\cdot \frac{1}{1+\g \f}\cdot \frac{1}{1+\g \f} \cdot \g \cdot \od{\f}{\ln(M^p)}.
\end{equation*}
Using \eqref{e:ma}, we find that
\begin{equation*}
(\e-1)(1+\g \f) M = \e + (\e-1)\g \f.
\end{equation*}
Moreover, by definition, the superelasticity $\s$ of the fairness function satisfies $\f\s = \odx{\f}{\ln(M^p)}$. Combining these three results, we obtain
\begin{equation}
-\odl{M}{M^p} = \frac{\g \f \s}{[\e+(\e-1)\g \f](1+\g \f)}.
\label{e:dmdmp}\end{equation}

Finally, combining \eqref{e:beta0} with \eqref{e:dmdmp} yields the cost passthrough
\begin{equation}
\b=\frac{1}{1+\frac{\g^{2}\f \s}{\bp{1+\g \f} \bs{\e+(\e-1) \g\f}}}.
\label{e:betaa}\end{equation} 
Since $\g>0$ (Definition~\ref{d:subproportional}), $\f>0$ (Lemma~\ref{l:phi}), and $\s>0$ (also Lemma~\ref{l:phi}), we infer that $\b\in(0,1)$.

\subsection{Proof of Corollary \ref{c:subproportional}}\label{a:corollary}

We apply the results of Proposition~\ref{p:subproportional} to a specific fairness function:
\begin{equation}
F(M^p) = 1- \t \cdot \bp{M^p - \frac{\e}{\e-1}}.
\label{e:fa}\end{equation}
We also assume that customers are acclimated, so $M^p = \e/(\e-1)$ and $F=1$.

\paragraph{Preliminary Results} The elasticity of the fairness function \eqref{e:fa} is
\begin{equation*}
\f = - \frac{M^p}{F}\cdot \od{F}{M^p} = \frac{M^p}{F}\cdot \t.
\end{equation*}
Accordingly, the superelasticity of the fairness function \eqref{e:fa} satisfies
\begin{equation*}
\s = \odl{\f}{M^p} = 1 - \odl{F}{M^p} = 1+\f.
\end{equation*}
When $M^p = \e/(\e-1)$ and $F=1$, the elasticity and superelasticity simplify to 
\begin{align}
\f &= \frac{\e\t }{\e-1}\label{e:phia}\\
\s &= 1+\frac{\e\t }{\e-1}.\label{e:sigmaa}
\end{align}

\paragraph{Markup} Combining \eqref{e:ma} with \eqref{e:phia}, we obtain the following markup:
\begin{equation*}
M = 1+\frac{1}{\e-1}\cdot \frac{1}{1+\g\e\t/(\e-1)} = 1+\frac{1}{(1+\g\t)\e-1}.
\end{equation*}
This expression shows that $M$ is lower when $\e$, $\g$, or $\t$ are higher.

\paragraph{Passthrough} Combining \eqref{e:betaa} with \eqref{e:phia} and \eqref{e:sigmaa}, we find that the cost passthrough $\b$ satisfies
\begin{align*}
1/\b = 1+\frac{\g^{2}\e\t \bs{(\e-1)+\e\t}}{(\e-1)\bs{(\e-1)+\g \e\t} \bp{\e+\g\e\t}}= 1+\frac{\g^{2}\t \bs{(1+\t)\e-1}}{(\e-1)\bs{(1+\g\t)\e-1} \bp{1+\g\t}}.
\end{align*}

Next we introduce the auxiliary function
\begin{equation}
\D(\g,\t,\e) = \frac{\g^{2}\t \bs{(1+\t)\e-1}}{(\e-1)\bs{(1+\g\t)\e-1} \bp{1+\g\t}},
\label{e:delta}\end{equation}
where $\g\in (0,1]$, $\t>0$, and $\e>1$. 

First, we divide numerator and denominator of $\D$  by $(\e-1)$:
\begin{equation*}
\D(\g,\t,\e) = \frac{\g^{2}\t\bs{1+\t\e/(\e-1)}}{\bs{(1+\g\t)\e-1} \bp{1+\g\t}}.
\end{equation*}
Since $\e/(\e-1)$ is decreasing in $\e$ and $(1+\g\t)\e-1$ is increasing in $\e$, $\D$ is decreasing in $\e$. As $\b = 1/(1+\D)$, we conclude that $\b$ is increasing in $\e$.

Next, we divide numerator and denominator of $\D$ in \eqref{e:delta} by $\t \bp{\e\t+\e-1}$:
\begin{equation*}
\D(\g,\t,\e) = \frac{\g^{2}}{(\e-1)\bp{\g+1/\t}\frac{\g\e\t+\e-1}{\e\t+\e-1}}.
\end{equation*}
First, $\g+1/\t$ is decreasing in $\t$. Second, as $\e>1$ and $\g\leq 1$,  $(\g\e\t+\e-1)/(\e\t+\e-1)$ is decreasing in $\t$. Hence, $\D$ is increasing in $\t$, and as $\b = 1/(1+\D)$, $\b$ is decreasing in $\t$.

Last, dividing numerator and denominator of $\D$ in \eqref{e:delta} by $\g^2$, we get
\begin{equation*}
\D(\g,\t,\e) = \frac{\t \bs{(1+\t)\e-1}}{(\e-1)\bs{\t\e+(\e-1)/\g} \bp{\t+1/\g}}.
\end{equation*}
The denominator is decreasing in $\g$, so $\D$ is increasing in $\g$. Since $\b = 1/(1+\D)$, we conclude that $\b$ is decreasing in $\g$.

\section{Derivations for the New Keynesian Model}\label{a:nk}

We derive the properties of the New Keynesian model with fairness presented in Section~\ref{s:nk}. In particular, we prove Lemmas \ref{l:monetary} and \ref{l:longrun}, Propositions \ref{p:monetary} and \ref{p:longrun}, and Corollary~\ref{c:longrun}.

\subsection{Household and Firm Problems}

We begin by solving the problems of households and firms.

\paragraph{Household $k$'s Problem} To solve household~$k$'s problem, we set up the Lagrangian:
\begin{align*}
\Lc_{k}& =\E[0]\sum_{t=0}^{\infty}\d^{t}  \bigg\{\ln{Z_{k}(t)}-\frac{N_{k}(t)^{1+\eta} }{1+\eta}\\
& + \Ac_{k}(t)  \bs{W_{k}(t)  N_{k}(t)+ B_{k}(t-1)+V_{k}(t)- Q(t)  B_{k}(t)-\int_{0}^{1} P_{j}(t)  Y_{jk}(t)\,dj}\\
&+\Bc_{k}(t)  \bs{N_{k}^{d}(t,W_{k}(t))-N_{k}(t)}\bigg\}.
\end{align*}
In the Lagrangian, $\Ac_{k}(t)$ is the Lagrange multiplier on the budget constraint in period~$t$; $\Bc_{k}(t)$ is the Lagrange multiplier on the labor-demand constraint in period~$t$; and $Z_{k}(t)$ is the fairness-adjusted consumption index: 
\begin{equation}
Z_{k}(t) = \bs{\int_{0}^{1} Z_{jk}(t)^{(\e-1)/\e}\,dj}^{\e/(\e-1)}.
\label{e:zk}\end{equation}
In the consumption index, $Z_{jk}(t)$ is the fairness-adjusted consumption of good~$j$:
\begin{equation}
Z_{jk}(t)= F_{j}(t) \cdot Y_{jk}(t).
\label{e:zjk}\end{equation}

\paragraph{First-Order Conditions with Respect to Consumption} We first compute the first-order conditions with respect to $Y_{jk}(t)$ for all goods $j\in[0,1]$: $\pdx{\Lc_{k}}{Y_{jk}(t)}=0$. From the definitions of $Z_{k}(t)$ and $Z_{jk}(t)$ given by \eqref{e:zk} and \eqref{e:zjk}, we find 
\begin{equation*}
\pd{Z_{jk}(t)}{Y_{jk}(t)}=F_{j}(t)\quad\text{and}\quad \pd{Z_{k}(t)}{Z_{jk}(t)}=\bs{\frac{Z_{jk}(t)}{Z_{k}(t)}}^{-1/\e} dj.
\end{equation*} 
Hence the first-order conditions imply that for all $j\in[0,1]$,
\begin{equation}
\bs{\frac{Z_{jk}(t)}{Z_{k}(t)}}^{-1/\e}  \frac{F_{j}(t)}{Z_{k}(t)}=\Ac_{k}(t)  P_{j}(t).
\label{e:focy}\end{equation} 

Taking \eqref{e:focy} to the power of $1-\e$ and shuffling terms, we then obtain
\begin{equation*}
\frac{1}{Z_{k}(t)^{1-\e}} \cdot \frac{1}{Z_{k}(t)^{(\e-1)/\e}}\cdot Z_{jk}(t)^{(\e-1)/\e}= \Ac_{k}(t)^{1-\e}  \bs{\frac{P_{j}(t)}{F_{j}(t)}}^{1-\e}.
\end{equation*} 
We integrate this equation over $j\in[0,1]$, use the definition of $Z_{k}(t)$ given by \eqref{e:zk}, and introduce the price index
\begin{equation}
X(t)=\bc{\int_0^1\bs{\frac{P_{j}(t)}{F_{j}(t)}}^{1-\e}\,dj}^{1/(1-\e)}.
\label{e:x}\end{equation}
We obtain the following:
\begin{equation*}
\frac{1}{Z_{k}(t)^{1-\e}}\cdot \frac{Z_{k}(t)^{(\e-1)/\e}}{Z_{k}(t)^{(\e-1)/\e}}= \Ac_{k}(t)^{1-\e} X(t)^{1-\e}.
\end{equation*}
From this equation we infer
\begin{equation}
\Ac_{k}(t) = \frac{1}{X(t) Z_{k}(t)}.
\label{e:ak}\end{equation}

Last, combining \eqref{e:focy} and \eqref{e:ak}, we find that the optimal fairness-adjusted consumption of good~$j$ by household~$k$ satisfies
\begin{equation*}
Z_{jk}(t) = Z_{k}(t) \bs{\frac{P_{j}(t)}{X(t)}}^{-\e}F_{j}(t)^{\e}.
\end{equation*}
As consumption and fairness-adjusted consumption of good~$j$ are related by $Y_{jk}(t)= Z_{jk}(t) / F_{j}(t)$, the optimal consumption of good~$j$ by household~$k$ satisfies
\begin{equation}
Y_{jk}(t)=  Z_{k}(t) \bs{\frac{P_{j}(t)}{X(t)}}^{-\e} F_{j}(t)^{\e-1}.
\label{e:yjk}\end{equation}
Integrating \eqref{e:yjk} over all households $k\in[0,1]$ yields the output of good~$j$:
\begin{equation*}
Y_{j}(t) = Z(t) \bs{\frac{P_{j}(t)}{X(t)}}^{-\e} F_{j}(t)^{\e-1}.
\end{equation*}
We note that the fairness factor $F_{j}(t)$ is a function of the perceived price markup, $F_{j}(t)=F_j(P_{j}(t)/C_{j}^{p}(t))$, and that the perceived marginal cost $C_{j}^{p}(t)$ follows the law of motion \eqref{e:cpj}. These observations allow us to obtain the demand for good~$j$:
\begin{equation*}
Y^{d}_{j}(t,P_{j}(t),C_{j}^{p}(t-1)) = Z(t) \bs{\frac{P_{j}(t)}{X(t)}}^{-\e} F_j\of{\bp{\frac{\e}{\e-1}}^{1-\g}\bs{\frac{P_{j}(t)}{C_{j}^{p}(t-1)}}^{\g}}^{\e-1}.
\end{equation*}

For future reference, the elasticities of the function $Y^{d}_{j}(t,P_{j}(t),C_{j}^{p}(t-1))$ are
\begin{align}
-\pdl{Y^{d}_{j}}{P_{j}}&=\e+(\e-1) \g \f_{j}(M^{p}_{j}(t)) \equiv E_{j}(M^{p}_{j}(t))\label{e:ydp}\\
\pdl{Y^{d}_{j}}{C_{j}^{p}}&=(\e-1) \g \f_{j}(M^{p}_{j}(t)) = E_{j}(M^{p}_{j}(t)) - \e,\label{e:ydc}
\end{align}
where $\f_{j} = -\odlx{F_{j}}{M^{p}_{j}}$ is the elasticity of the fairness function. The function $E_{j}$ gives the price elasticity of the demand for good~$j$.

Moreover, using \eqref{e:yjk} and the definition of the price index $X$ given by \eqref{e:x}, we find that
\begin{equation*}
\int_0^1 P_{j} Y_{jk}\,dj = X^{\e}  Z_{k} \int_0^1 \bp{\frac{P_{j}}{F_{j}}}^{1-\e}  \,dj = X Z_{k}.
\end{equation*}
This means that when households optimally allocate their consumption expenditures across goods, the price of one unit of fairness-adjusted consumption index is $X$.

\paragraph{First-Order Condition with Respect to Bonds} The first-order condition with respect to $B_{k}(t)$ is $\pdx{\Lc_{k}}{B_{k}(t)}=0$, which gives
\begin{equation*}
Q(t)  \Ac_{k}(t) = \d  \E[t]{\Ac_{k}(t+1)}.
\end{equation*} 
Using \eqref{e:ak}, we obtain household~$k$'s consumption Euler equation:
\begin{equation}
Q(t)=\d \E[t]{\frac{X(t) Z_{k}(t)}{X(t+1) Z_{k}(t+1)}}.
\label{e:eulerx}\end{equation}
This equation governs how the household smooths fairness-adjusted consumption over time.

\paragraph{Firm $j$'s Problem} Since the wages set by households depend on firms' labor demands, we turn to the firms' problems before finishing the households' problems. To solve firm~$j$'s problem, we set up the Lagrangian:
\begin{align*}
\Lc_{j}& =\E[0]\sum_{t=0}^{\infty}\G(t) \bigg\lbrace P_{j}(t)  Y_{j}(t)-\int_{0}^{1}W_{k}(t)  N_{jk}(t)\,dk \\
& + \Hc_{j}(t)  \bs{Y^{d}_{j}(t,P_{j}(t),C^{p}_{j}(t-1))-Y_{j}(t)} + \Jc_{j}(t)  \bs{A_{j}(t)  N_{j}(t)^{\a}-Y_{j}(t)}\\
&+\Kc_{j}(t)  \bs{C^{p}_{j}(t-1)^{\g}    \bs{\frac{\e-1}{\e} P_{j}(t)}^{1-\g}-C^{p}_{j}(t)}\bigg\rbrace.
\end{align*}
In the Lagrangian, $\Hc_{j}(t)$ is the Lagrange multiplier on the demand constraint in period~$t$; $\Jc_{j}(t)$ is the Lagrange multiplier on the production constraint in period~$t$; $\Kc_{j}(t)$ is the Lagrange multiplier on the law of motion of the perceived marginal cost in period~$t$; and $N_{j}(t)$ is the employment index:
\begin{equation}
N_{j}(t)= \bs{\int_{0}^1 N_{jk}(t)^{(\n-1)/\n}\,dk}^{\n/(\n-1)}.
\label{e:nj}\end{equation}

\paragraph{First-Order Conditions with Respect to Employment}  We compute the first-order conditions with respect to $N_{jk}(t)$ for all labor services $k\in[0,1]$: $\pdx{\Lc_{j}}{N_{jk}(t)}=0$. From the definition of $N_{j}(t)$ given by \eqref{e:nj}, we know that 
\begin{equation*}
\pd{N_{j}(t)}{N_{jk}(t)}=\bs{\frac{N_{jk}(t)}{N_{j}(t)}}^{-1/\n} dk.
\end{equation*}
Hence the first-order conditions imply that for all $k\in[0,1]$,
\begin{equation}
W_{k}(t)=  \a \Jc_{j}(t) A_{j}(t) N_{j}(t)^{\a-1}\bs{\frac{N_{jk}(t)}{N_{j}(t)}}^{-1/\n}.
\label{e:wk}\end{equation}

Toward deriving firm~$j$'s labor demand, we introduce the wage index
\begin{equation}
W(t) = \bs{\int_{0}^1 W_{k}(t)^{1-\n}\,dk}^{1/(1-\n)}.
\label{e:w}\end{equation}
Taking \eqref{e:wk} to the power of $1-\n$, we obtain
\begin{equation*}
 W_{k}(t)^{1-\n} = \bs{\a \Jc_{j}(t)  A_{j}(t)  N_{j}(t)^{\a-1}}^{1-\n} \frac{1}{N_{j}(t)^{(\n-1)/\n}} N_{jk}(t)^{(\n-1)/\n}.
\end{equation*}
Integrating this condition over $k\in[0,1]$ and using the definitions of $N_{j}$ and $W$  given by \eqref{e:nj} and \eqref{e:w}, we find
\begin{equation*}
W(t)^{1-\n} = \bs{\a \Jc_{j}(t)  A_{j}(t)  N_{j}(t)^{\a-1}}^{1-\n} \frac{N_{j}(t)^{(\n-1)/\n}}{N_{j}(t)^{(\n-1)/\n}}.
\end{equation*}
From this equation we infer
\begin{equation}
W(t) = \a \Jc_{j}(t) A_{j}(t)  N_{j}(t)^{\a-1}.
\label{e:jj}\end{equation}

Last, we combine \eqref{e:wk} and \eqref{e:jj} to determine the quantity of labor that firm~$j$ hires from household~$k$:
\begin{equation}
N_{jk}(t)=N_{j}(t) \bs{\frac{W_{k}(t)}{W(t)}}^{-\n}.
\label{e:njk}\end{equation}
Integrating \eqref{e:njk} over all firms~$j\in[0,1]$ yields the demand for labor service~$k$:
\begin{equation}
N_{k}^{d}(t,W_{k}(t)) = N(t) \bs{\frac{W_{k}(t)}{W(t)}}^{-\n},
\label{e:nk}\end{equation}
where $N(t) = \int_0^1 N_{j}(t)\,dj$ is aggregate employment.

Moreover, \eqref{e:w} and \eqref{e:njk} imply that
\begin{equation*}
\int_0^1 W_{k} N_{jk}\,dk = W^{\n}  N_{j} \int_0^1 W_{k}^{1-\n} \,dk = W N_{j}.
\end{equation*}
This means that when firms optimally allocate their wage bill across labor services, the cost of one unit of labor index is $W$.

\paragraph{First-Order Conditions with Respect to Labor and Wage} We now finish solving household~$k$'s problem using labor demand~\eqref{e:nk}. The first-order conditions with respect to $N_{k}(t)$ and $W_{k}(t)$ are $\pdx{\Lc_{k}}{N_{k}(t)}=0$ and $\pdx{\Lc_{k}}{W_{k}(t)}=0$; they yield
\begin{align}
N_{k}(t)^{\eta}&=\Ac_{k}(t)  W_{k}(t)-\Bc_{k}(t)\label{e:focnk}\\
\Ac_{k}(t) N_{k}(t)&=-\Bc_{k}(t)  \od{N_{k}^{d}}{W_{k}}.\label{e:focwk}
\end{align}
Since the elasticity of $N_{k}^{d}$ with respect to $W_{k}$ is $-\n$, we infer from \eqref{e:focwk} that
\begin{equation}
\Ac_{k}(t) W_{k}(t) = \Bc_{k}(t) \n.
\label{e:akwk}\end{equation}
Plugging this result into \eqref{e:focnk}, we obtain
\begin{equation*}
\Bc_{k}(t) = \frac{N_{k}(t)^{\eta}}{\n-1}.
\end{equation*}
Combining this result with \eqref{e:akwk} then yields
\begin{equation*}
W_{k}(t) = \frac{\n}{\n-1} \cdot \frac{N_{k}(t)^{\eta}}{\Ac_{k}(t)}. 
\end{equation*}
Finally, by merging this equation with \eqref{e:ak}, we find that household~$k$ sets its wage rate at
\begin{equation}
\frac{W_{k}(t)}{X(t)}=\frac{\n}{\n-1} N_{k}(t)^{\eta} Z_{k}(t).
\label{e:wkx}\end{equation}
This equation shows that households set their real wage at a markup of $\n/(\n-1)>1$ over the marginal rate of substitution between leisure and consumption. 

\paragraph{First-Order Condition with Respect to Output} We then finish solving firm~$j$'s problem. The first-order condition with respect to $Y_{j}(t)$ is $\pdx{\Lc_{j}}{Y_{j}(t)}=0$, which gives 
\begin{equation*}
P_{j}(t)=\Hc_{j}(t)+\Jc_{j}(t).
\end{equation*}
Using the value of $\Jc_{j}(t)$ given by \eqref{e:jj}, we then obtain 
\begin{equation}
\Hc_{j}(t)=P_{j}(t) \bs{1-\frac{W(t)/P_{j}(t)}{ \a A_{j}(t) N_{j}(t)^{\a-1}}}.
\label{e:hj1}\end{equation}
Firm~$j$'s nominal marginal cost is the nominal wage divided by the marginal product of labor:
\begin{equation}
C_{j}(t)= \frac{W(t)}{\a A_{j}(t) N_{j}(t)^{\a-1}}.
\label{e:cj}\end{equation}
Hence the first-order condition \eqref{e:hj1} can be written
\begin{equation}
\Hc_{j}(t)=P_{j}(t) \bs{1-\frac{C_{j}(t)}{P_{j}(t)}}.
\label{e:hj2}\end{equation}
Given that firm~$j$'s markup is $M_j(t)=P_j(t)/C_j(t)$, we rewrite this equation as
\begin{equation}
\frac{\Hc_{j}(t)}{P_{j}(t)} = \frac{M_{j}(t)-1}{M_{j}(t)}.
\label{e:hj}\end{equation}

\paragraph{First-Order Condition with Respect to Price} The first-order condition of firm~$j$'s problem with respect to $P_{j}(t)$ is $\pdx{\Lc_{j}}{P_{j}(t)}=0$. It yields
\begin{equation}
0 = Y_{j}(t) +\Hc_{j}(t)\pd{Y_{j}^{d}}{P_{j}} +(1-\g)\Kc_{j}(t) \frac{C^{p}_{j}(t)}{P_{j}(t)}.
\label{e:focpj}\end{equation}
We divide this condition by $Y_{j}(t)$ and insert the price elasticity of the demand for good~$j$, $E_{j}(M^{p}_{j}(t)) = - \pdlx{Y_{j}^{d}}{P_j}$, as well as the perceived price markup for good~$j$, $M^{p}_{j}(t) = P_{j}(t)/C^{p}_{j}(t)$. We obtain
\begin{equation*}
0 = 1-\frac{\Hc_{j}(t)E_{j}(M^{p}_{j}(t))}{P_{j}(t)}+(1-\g)\frac{\Kc_{j}(t)}{Y_{j}(t) M^{p}_{j}(t)}.
\end{equation*}
Using the value of $\Hc_{j}(t)$ given by \eqref{e:hj}, we finally obtain
\begin{equation}
(1-\g)\frac{\Kc_j(t)}{Y_{j}(t)M^{p}_{j}(t)} = \frac{M_{j}(t)-1}{M_{j}(t)} E_{j}(M^{p}_{j}(t))-1.
\label{e:ejmj}\end{equation}

\paragraph{First-Order Condition with Respect to Perceived Marginal Cost} The first-order condition of firm~$j$'s problem with respect to $C^{p}_{j}(t)$ is $\pdx{\Lc_{j}}{C^p_{j}(t)}=0$. It gives
\begin{equation*}
0 =\E[t]{\frac{\G(t+1)}{\G(t)} \Hc_{j}(t+1)  \pd{Y_{j}^{d}}{C^{p}_{j}}} + \g  \E[t]{ \frac{\G(t+1)}{\G(t)}  \Kc_{j}(t+1)   \frac{C^{p}_{j}(t+1)}{C^{p}_{j}(t)}} -\Kc_{j}(t).
\end{equation*}
And using the elasticity given by \eqref{e:ydc}, we find
\begin{equation*}
\Kc_{j}(t)=\E[t]{\frac{\G(t+1)}{\G(t)}\bc{\Hc_{j}(t+1)\frac{Y_{j}(t+1)}{C_{j}^{p}(t)} [E_{j}(M^{p}_{j}(t+1))-\e]+ \g \Kc_{j}(t+1) \frac{C^{p}_{j}(t+1)}{C_{j}^{p}(t)}}}.
\end{equation*}
We modify this equation in two steps: first, we multiply it by $C_j^p(t)/[Y_j(t)P_j(t)]$; second, we insert the perceived price markups $M_j^p(t) = P_j(t)/C_j^{p}(t)$ and $M_j^p(t+1) = P_j(t+1)/C_j^{p}(t+1)$. We get
\begin{equation*}\medmuskip=2mu
\frac{\Kc_{j}(t)M_j^p(t)}{Y_j(t)}=\E[t]{\frac{\G(t+1)Y_j(t+1)P_j(t+1)}{\G(t)Y_j(t)P_j(t)}\bc{\frac{\Hc_{j}(t+1)}{P_j(t+1)}[E_{j}(M^{p}_{j}(t+1))-\e]+ \g\frac{\Kc_{j}(t+1) M^{p}_{j}(t+1)}{Y_{j}(t+1)}}}.
\end{equation*}
Last, we multiply the equation by $(1-\g)$; and we eliminate $\Hc_j(t+1)$ using \eqref{e:hj} and $\Kc_j(t)$ and $\Kc_j(t+1)$ using \eqref{e:ejmj}. We obtain firm~$j$'s pricing equation, which links its markup to its perceived markup:
\begin{align}
&\frac{M_{j}(t)-1}{M_{j}(t)} E_{j}(M^{p}_{j}(t))= \label{e:emj}\\
&1+\E[t]{\frac{\G(t+1)Y_j(t+1)P_j(t+1)}{\G(t)Y_j(t)P_j(t)}\bc{\frac{M_{j}(t+1)-1}{M_{j}(t+1)}[E_{j}(M^{p}_{j}(t+1))-(1-\g)\e]-\g}}.\nonumber
\end{align}

\subsection{Equilibrium}

We present the equilibrium of the model. Because all households and firms face the same conditions, they all behave the same in equilibrium, so we drop the subscripts~$j$ and~$k$ on all variables.

The equilibrium can be described by seven variables: output $Y(t)$, employment $N(t)$, the price level $P(t)$, the wage $W(t)$, the bond price $Q(t)$, the price markup $M(t)$, and the perceived price markup $M^p(t)$. Seven equations determine these seven variables.

The first equation is the monetary-policy rule, given by \eqref{e:taylor}. This equation links the nominal interest rate, $i(t)$, to the inflation rate, $\pi(t)$. By definition, however, the nominal interest rate is determined by the bond price and the inflation rate by the price level:
\begin{equation}
i(t) = \ln{\frac{1}{Q(t)}}, \qquad  \pi(t) = \ln{\frac{P(t)}{P(t-1)}}.
\label{e:definition}\end{equation}  
Hence the monetary-policy rule links bond price to price level. 

The second equation is the production function, which is directly obtained from \eqref{e:yj}:
\begin{equation}
Y(t) = A(t) N(t)^{\a}.
\label{e:y}\end{equation}

The third equation is the usual consumption Euler equation, obtained by simplifying \eqref{e:eulerx}. By symmetry $X(t)=P(t)/F(t)$ and $Z_{k}(t)=F(t) Y(t)$, so \eqref{e:eulerx} gives
\begin{equation}
Q(t)=\d \E[t]{\frac{P(t) Y(t)}{P(t+1) Y(t+1)}}.
\label{e:euler}\end{equation}

The fourth equation is the usual expression for the real wage, obtained by simplifying \eqref{e:wkx}. Once again, by symmetry $X(t)=P(t)/F(t)$ and $Z_{k}(t)=F(t) Y(t)$, so \eqref{e:wkx} yields
\begin{equation*}
\frac{W(t)}{P(t)}=\frac{\n}{\n-1}  N(t)^{\eta}  Y(t).
\end{equation*}
Combining this equation with \eqref{e:y}, we express the real wage as a function of employment:
\begin{equation}
\frac{W(t)}{P(t)}=\frac{\n}{\n-1}  A(t) N(t)^{\eta+\a}.
\label{e:wp}\end{equation}

The fifth equation is the standard link between price markup and employment, which is obtained from the definition of the price markup. In a symmetric economy the price markup is just the inverse of the real marginal cost: $M(t) = P(t)/C(t)$. Combining the expression of the nominal marginal cost given by \eqref{e:cj} with the value of the real wage given by \eqref{e:wp}, we infer the real marginal cost:
\begin{equation*}
\frac{C(t)}{P(t)} = \frac{\n}{(\n-1) \a} N(t)^{1+\eta}.
\end{equation*}
Since the price markup is the inverse of the real marginal cost, we find
\begin{equation}
N(t)=\bs{\frac{(\n-1) \a }{\n} \cdot  \frac{1}{M(t)}}^{1/(1+\eta)}.
\label{e:na}\end{equation}

The sixth equation is a pricing equation, which is obtained by simplifying \eqref{e:emj}. In equilibrium the stochastic discount factor is given by
\begin{equation*}
\G(t)= \d^{t}\cdot \frac{X(0)Z(0)}{X(t) Z(t)}. 
\end{equation*}
Since by symmetry $Z(t)=F(t) Y(t)$ and $X(t)=P(t)/F(t)$, we have
\begin{equation*}
\frac{\G(t+1)}{\G(t)}=\d \cdot\frac{X(t)}{X(t+1)}\cdot \frac{Z(t)}{Z(t+1)}=\d \cdot\frac{P(t)}{P(t+1)}\cdot \frac{Y(t)}{Y(t+1)}.
\end{equation*}
Hence, \eqref{e:emj} simplifies to
\begin{equation}
\frac{M(t)-1}{M(t)} E(M^p(t)) = 1-\d\g + \d\E[t]{\frac{M(t+1)-1}{M(t+1)}\bs{E(M^p(t+1))-(1-\g) \e}}.
\label{e:ema}\end{equation}
This pricing equation shows the dynamic relationship between actual and perceived price markups. Unlike the other equilibrium conditions---which are the same as in the textbook model---the pricing equation is unique to the model with fairness.

The seventh and final equation is the law of motion of the perceived price markup. It derives from the law of motion of the perceived marginal cost, given by \eqref{e:cpj}. Since $M^p(t)=P(t)/C^p(t)$, \eqref{e:cpj} implies
\begin{equation*}
M^p(t) = \bs{\frac{P(t)}{(\e-1)P(t)/\e}}^{1-\g} \bs{\frac{P(t)}{C^p(t-1)}}^{\g} = \bp{\frac{\e}{\e-1}}^{1-\g} \bs{\frac{P(t)}{P(t-1)}}^{\g} \bs{\frac{P(t-1)}{C^p(t-1)}}^{\g}.
\end{equation*}
Hence the perceived price markup satisfies
\begin{equation}
M^p(t) = \bp{\frac{\e}{\e-1}}^{1-\g} \bs{\frac{P(t)}{P(t-1)}}^{\g}\bs{M^p(t-1)}^{\g}.
\label{e:mpt}\end{equation}

\subsection{Steady-State Equilibrium}\label{a:steadystate}

We now apply the equilibrium conditions to a steady-state environment, in which all real variables are constant and all nominal variables grow at the inflation rate, $\ol{\pi}$. 
We use these steady-state conditions to prove Lemma~\ref{l:longrun}, Proposition~\ref{p:longrun}, and Corollary~\ref{c:longrun}. We also use these conditions to compute the long-run Phillips curves that are displayed in Figure \ref{f:longrun}.

We describe the steady-state equilibrium by six variables: output $\ol{Y}$, employment $\ol{N}$, inflation $\ol{\pi}$, nominal interest rate $\ol{i}$, price markup $\ol{M}$, and perceived price markup $\ol{M^p}$. These six variables are governed by six equations.

\paragraph{Steady-State Equilibrium Conditions} First, in steady state the consumption Euler equation \eqref{e:euler} gives 
\begin{equation*}
\ol{Q}=\d \cdot \frac{P(t)}{P(t+1)}. 
\end{equation*}
Taking the logarithm of this equation and using \eqref{e:definition}, we obtain 
\begin{equation*}
\ol{i} = \r + \ol{\pi},
\end{equation*}
 where $\r \equiv -\ln(\d)$ is the discount rate. Equivalently, the steady-state real interest rate $\ol{r} \equiv \ol{i}- \ol{\pi}$ equals the discount rate $\r$.

Second, in steady state the monetary-policy rule \eqref{e:taylor} implies that $\ol{r} = \ol{i_{0}} + (\p-1) \ol{\pi}$. Since $\ol{r} = \r$, the steady-state inflation rate is
\begin{equation*}	
\ol{\pi}=\frac{\r-\ol{i_{0}}}{\p-1}.
\end{equation*}

Third, in steady state the law of motion of the perceived price markup \eqref{e:mpt} implies that
\begin{equation*}
(\ol{M^p})^{1-\g}= \bp{\frac{\e}{\e-1}}^{1-\g} \bs{\frac{P(t)}{P(t-1)}}^{\g} .
\end{equation*}
Taking this expression to the power of $1/(1-\g)$, and noting that in steady state $P(t)/P(t-1)=\exp(\ol{\pi})$, we find that the steady-state perceived price markup is
\begin{equation}
\ol{M^{p}} = \frac{\e}{\e-1} \exp{\frac{\g}{1-\g} \ol{\pi}}.
\label{e:mpss}\end{equation}

Fourth, in steady state the pricing equation \eqref{e:ema} implies that
\begin{equation}
0 =  1-\d\g - \frac{\ol{M}-1}{\ol{M}} E(\ol{M^p}) + \d \frac{\ol{M}-1}{\ol{M}} \bs{E(\ol{M^p})-(1-\g) \e}.
\label{e:emss1}\end{equation}
Shuffling this expression, we obtain the following:
\begin{align}
0 &= (1-\d\g) \ol{M} - (\ol{M}-1) E(\ol{M^p}) + \d (\ol{M}-1) \bs{E(\ol{M^p})-(1-\g) \e}\nonumber\\
0 &= \bs{1-\d\g - (1-\d)E(\ol{M^p})- \d(1-\g)\e} \ol{M} + (1-\d)E(\ol{M^p})+\d(1-\g)\e\nonumber\\
\ol{M} &=\frac{(1-\d)E(\ol{M^p})+\d (1-\g)\e}{(1-\d)E(\ol{M^p})+\d(1-\g)\e-(1-\d\g)}\nonumber\\
\ol{M} &= 1+ \frac{(1-\d\g)}{(1-\d)E(\ol{M^p}) +(\d-\d\g)\e  - (1-\d\g)}.\label{e:emss2}
\end{align}
In addition, \eqref{e:ydp} shows that in steady state the price elasticity of demand is $E(\ol{M^p})=\e+(\e-1) \g \f(\ol{M^p})$. Using this expression, we rewrite the denominator of the fraction in \eqref{e:emss2} as
\begin{equation*}
(1-\d) \e + (1-\d)(\e-1)\g\f(\ol{M^p}) +(\d-\d\g)\e - (1-\d\g) = (\e-1) \bs{(1-\d\g) + (1-\d) \g \f(\ol{M^p})}.
\end{equation*}
Plugging this result back into \eqref{e:emss2}, we obtain the steady-state price markup:
\begin{equation}
\ol{M} = 1+ \frac{1}{\e-1}\cdot \frac{1}{1 + \frac{(1-\d)\g}{1-\d\g} \f(\ol{M^p})}.
\label{e:mssa}\end{equation}

Fifth, we apply the markup-employment relation \eqref{e:na} to the steady state to express employment:
\begin{equation}
\ol{N} = \bs{\frac{(\n-1) \a}{\n}\cdot \frac{1}{\ol{M}}}^{1/(1+\eta)}.
\label{e:nssa}\end{equation}

Sixth, we apply the production function \eqref{e:y} to the steady state to express output:
\begin{equation*}
\ol{Y} = \ol{A}\cdot\ol{N}^{\a}.
\end{equation*}

\paragraph{Proof of Lemma~\ref{l:longrun}} The expression for the steady-state perceived price markup $\ol{M^p}$ comes from \eqref{e:mpss}. The expression for the steady-state fairness factor $\ol{F} = F(\ol{M^p})$ follows from combining \eqref{e:facclim} with \eqref{e:mf}. Last, the expression for the steady-state elasticity of the fairness function 
\begin{equation*}
\ol{\f} = \f(\ol{M^p}) = -F'(\ol{M^p})\cdot \frac{\ol{M^p}}{F(\ol{M^p})}
\end{equation*} 
comes from noting that with the fairness function \eqref{e:facclim}, $F'(M^p) = -\t$. The properties that $\ol{M^p}$ and $\ol{\f}$ are strictly increasing in $\ol{\pi}$, and that $\ol{F}$ is weakly decreasing in $\ol{\pi}$, follow from the assumptions that $\e>1$, $\g\in(0,1)$, $\t>0$, and $1-\c \geq 0$.

\paragraph{Proof of Proposition~\ref{p:longrun}} The expressions for the steady-state price markup $\ol{M}$ and steady-state employment $\ol{N}$ come from \eqref{e:mssa} and \eqref{e:nssa}. Since $\d<1$,  $\g\in(0,1)$, and $\ol{\f}>0$ is strictly increasing in $\ol{\pi}$ (Lemma~\ref{l:longrun}), it follows that $\ol{M}$ is strictly decreasing in $\ol{\pi}$. And since $\a>0$, $\n>1$, $\eta>0$, and $\ol{M}>0$ is strictly decreasing in $\ol{\pi}$, it follows that $\ol{N}$ is strictly increasing in $\ol{\pi}$.

\paragraph{Proof of Corollary~\ref{c:longrun}} First, the expressions for the steady-state perceived price markup, $\ol{M^p}$, steady-state fairness factor, $\ol{F}$, and steady-state elasticity of the fairness function, $\ol{\f}$, in Lemma~\ref{l:longrun}indicate that around the zero-inflation steady state, 
\begin{equation}
\ol{M^p} = \frac{\e}{\e-1},\qquad \ol{F}=1,\qquad \ol{\f}=\frac{\t\e}{\e-1}.
\label{e:longrun}\end{equation}
These expressions also show that 
\begin{align*}
\od{\ln(\ol{M^p})}{\ol{\pi}} & = \frac{\g}{1-\g} \\
\od{\ln(\ol{F})}{\ol{\pi}} & = - \t \cdot (1-\c) \cdot \frac{\ol{M^p}}{\ol{F}} \cdot \od{\ln(\ol{M^p})}{\ol{\pi}}\\
\od{\ln(\ol{\f})}{\ol{\pi}} & =\od{\ln(\ol{M^p})}{\ol{\pi}} - \od{\ln(\ol{F})}{\ol{\pi}}.
\end{align*}
Hence, around the zero-inflation steady state, we have
\begin{equation*}
\od{\ln(\ol{F})}{\ol{\pi}} = - (1-\c) \cdot \frac{\t\e}{\e-1} \cdot \frac{\g}{1-\g}
\end{equation*}
and
\begin{equation}
\od{\ln(\ol{\f})}{\ol{\pi}} =  \frac{\g}{1-\g} \bs{1 + (1-\c) \cdot \frac{\t\e}{\e-1}}.
\label{e:dphi}\end{equation}

Second, the expression of the steady-state price markup $\ol{M}$ in Proposition~\ref{p:longrun} yields
\begin{equation}
\odl{\ol{M}}{\ol{\f}} = \frac{\ol{\f}}{\ol{M}}\cdot \od{\ol{M}}{\ol{\f}} = \frac{\ol{\f}}{\ol{M}}\cdot \frac{1}{\e-1}\cdot \frac{-1}{\bs{1+\frac{(1-\d)\g}{1-\d\g}\ol{\f}}^2}\cdot \frac{(1-\d)\g}{1-\d\g},
\label{e:dm}\end{equation}
and 
\begin{equation}
(\e-1) \bs{1+\frac{(1-\d)\g}{1-\d\g}\ol{\f}} \ol{M} = 1 + (\e-1) \bs{1+\frac{(1-\d)\g}{1-\d\g}\ol{\f}} = \e + (\e-1) \frac{(1-\d)\g}{1-\d\g}\ol{\f}.
\label{e:mssa2}\end{equation}
Combining \eqref{e:dm} and \eqref{e:mssa2}, we obtain
\begin{equation*}
-\odl{\ol{M}}{\ol{\f}} = \frac{(1-\d)\g}{1-\d\g} \cdot \ol{\f} \cdot \frac{1}{\e + (\e-1) \frac{(1-\d)\g}{1-\d\g}\ol{\f}}\cdot \frac{1}{1+\frac{(1-\d)\g}{1-\d\g}\ol{\f}}.
\end{equation*}
Hence, using \eqref{e:longrun}, we find that around the zero-inflation steady state,
\begin{equation}
-\odl{\ol{M}}{\ol{\f}} = \frac{(1-\d)\g}{1-\d\g} \cdot \frac{\t}{1+\frac{(1-\d)\g}{1-\d\g}\t}\cdot \frac{1}{\bs{1+\frac{(1-\d)\g}{1-\d\g}\t}\e-1}.
\label{e:dm2}\end{equation}

Third, from the expression of steady-state employment $\ol{N}$ in Proposition~\ref{p:longrun}, we learn that
\begin{equation*}
\od{\ln(\ol{N})}{\ol{\pi}} = \frac{-1}{1+\eta}\cdot\odl{\ol{M}}{\ol{\f}}\cdot\od{\ln(\ol{\f})}{\ol{\pi}}.
\end{equation*}
Using \eqref{e:dphi} and \eqref{e:dm2}, we therefore find
\begin{equation*}
\od{\ln(\ol{N})}{\ol{\pi}} = \frac{1-\d}{1+\eta}\cdot\frac{\t}{1+\frac{(1-\d)\g}{1-\d\g}\t}\cdot \frac{1}{\bs{1+\frac{(1-\d)\g}{1-\d\g}\t}\e-1}\cdot \frac{\g^2}{(1-\g)(1-\d\g)} \cdot\frac{\bs{1+(1-\c)\t}\e-1}{\e-1}.
\end{equation*}
Inverting this equation, we obtain the slope of the long-run Phillips curve:
\begin{equation}
\od{\ol{\pi}}{\ln(\ol{N})} = \frac{1+\eta}{1-\d}\cdot \frac{(1-\g)(1-\d\g)}{\g^2}\cdot \frac{\e-1}{\t} \cdot \frac{\bs{1+\frac{(1-\d)\g}{1-\d\g}\t}\bs{\bp{1+\frac{(1-\d)\g}{1-\d\g}\t}\e-1}}{\bs{1+(1-\c)\t}\e-1}.
\label{e:dpi}\end{equation}

We now derive comparative statics on the slope of the long-run Phillips curve. We repeatedly use the assumptions that $\d\in(0,1)$, $\eta>0$, $\g\in(0,1)$, $\t>0$, $\e>1$, and $\c\in[0,1]$. 

First, $\e$ influences the slope of the long-run Phillips curve through
\begin{equation*}
(\e-1) \frac{\bs{1+\frac{(1-\d)\g}{1-\d\g}\t}\e-1}{\bs{1+(1-\c)\t}\e-1},
\end{equation*}
which can be rewritten as
\begin{equation*}
\frac{\bs{1+\frac{(1-\d)\g}{1-\d\g}\t}\e-1}{(1-\c)\t\frac{\e}{\e-1}+1}.
\end{equation*}
Since $\e/(\e-1)$ is decreasing in $\e$ and 
\begin{equation*}
\bs{1+\frac{(1-\d)\g}{1-\d\g}\t}\e-1
\end{equation*}
 is increasing in $\e$, the slope of the long-run Phillips curve is increasing in $\e$.

Second, since $\c$ appears only once in \eqref{e:dpi}, it is clear that the slope of the long-run Phillips curve is increasing in $\c$.

Third, $\t$ influences the slope of the long-run Phillips curve through
\begin{equation*}
\frac{1}{\t} \cdot \frac{\bs{1+\frac{(1-\d)\g}{1-\d\g}\t}\bs{\bp{1+\frac{(1-\d)\g}{1-\d\g}\t}\e-1}}{\bs{1+(1-\c)\t}\e-1},
\end{equation*}
which can be rewritten as
\begin{equation*}
\X(\t) = \frac{\bs{1+\frac{(1-\d)\g}{1-\d\g}\t}\bs{\frac{(1-\d)\g}{1-\d\g}\t+\frac{\e-1}{\e}}}{\t\bs{(1-\c)\t +\frac{\e-1}{\e}}}.
\end{equation*}
The function $\X(\t)$ is a quadratic-quadratic rational function, whose behavior can be determined from its asymptotes and zeros. The function has two vertical asymptotes, at 
\begin{equation*}
\t = -\frac{\e-1}{(1-\c)\e}<0 \quad \text{and}\quad \t=0,
\end{equation*}
 and a horizontal asymptote, at 
\begin{equation}
\X = \bs{\frac{(1-\d)\g}{1-\d\g}}^2 \cdot \frac{1}{1-\c}>0.
\label{e:Xi}\end{equation}
Moreover, both zeros of $\X(\t)$ are negative, at
\begin{equation*}
\t = -\frac{1-\d\g}{(1-\d)\g}<0 \quad \text{and}\quad \t= -\frac{\e-1}{\e}\cdot\frac{1-\d\g}{(1-\d)\g}<0.
\end{equation*}
Hence, the function $\X(\t)$ cannot cross the $x$-axis when $\t>0$. This means that it must approach its positive horizontal asymptote from above, decreasing from $+\infty$ when $\t\to 0^+$ toward the horizontal asymptote \eqref{e:Xi} when $\t\to +\infty$. Thus, $\X(\t)$ is decreasing in $\t>0$, and so is the slope of the long-run Phillips curve.

Fourth, $\g$ influences the slope of the long-run Phillips curve through
\begin{equation*}
\frac{(1-\g)(1-\d\g)}{\g^2} \bs{1+\frac{(1-\d)\g}{1-\d\g}\t}\bs{\e-1+\frac{(1-\d)\g}{1-\d\g}\t\e},
\end{equation*}
which can be rewritten as
\begin{equation*}
\bs{\frac{1-\d\g}{\g}+(1-\d)\t}\bs{\frac{(\e-1)(1-\g)}{\g}+\frac{1-\g}{1-\d\g}(1-\d)\t\e}.
\end{equation*}
First, $(1-\d\g)/\g$ and $(1-\g)/\g$ are decreasing in $\g$. Second, since $\d<1$, $(1-\g)/(1-\d\g)$ is decreasing in $\g$. Thus, the slope of the long-run Phillips curve is decreasing in $\g$.

\subsection{Log-Linearized Equilibrium}\label{a:loglinear}

We log-linearize the equilibrium conditions around a steady state. We then use these log-linearized conditions to prove Lemma~\ref{l:monetary} and Proposition~\ref{p:monetary}. We also use these conditions to compute the impulse responses to monetary and technology shocks that are presented in Figures \ref{f:monetary} and~\ref{f:technology}.

We describe the log-linearized equilibrium through six variables. The first four are the log-deviations from steady state of output, employment, price markup, and perceived price markup: $\wh{y}(t)$, $\wh{n}(t)$, $\wh{m}(t)$, and $\wh{m^p}(t)$. The final two are the deviations from steady state of the nominal interest rate and inflation rate: $\wh{i}(t)$ and $\wh{\pi}(t)$. These six variables are governed by six linear equations.

\paragraph{Log-Linear Equilibrium Conditions} Several equilibrium conditions take a log-linear form, so they can immediately be log-linearized. The first is the monetary-policy rule \eqref{e:taylor}, which implies
\begin{equation}
\wh{i}(t)=\wh{i_0}(t)+\p \wh{\pi}(t). 
\label{e:ihat}\end{equation}
The second is the production function \eqref{e:y}, which gives
\begin{equation}
\wh{y}(t)= \wh{a}(t) + \a  \wh{n}(t).
\label{e:yhat}\end{equation} 
The third is the markup-employment relation \eqref{e:na}, which yields
\begin{equation}
\wh{m}(t)=-(1+\eta) \wh{n}(t). 
\label{e:mhat}\end{equation} 
The fourth is the law of motion for the perceived price markup \eqref{e:mpt}, which gives
\begin{equation}
\wh{m^p}(t) = \g  \bs{\wh{\pi}(t) + \wh{m^p}(t-1)}.
\label{e:mphata}\end{equation}

\paragraph{IS Equation} The fifth equation is the IS equation, which is based on the consumption Euler equation~\eqref{e:euler}. We start by computing a log-linear approximation of \eqref{e:euler}, as in \ct[pp.~35--36]{G08}:
\begin{equation*}
\ln(Y(t))=\E[t]{\ln(Y(t+1))} + \E[t]{\pi(t+1)} + \r - i(t),
\end{equation*} 
where $\r = -\ln(\d)$ is the discount rate. Subtracting the steady-state values of both sides yields
\begin{equation*}
\wh{y}(t)=\E[t]{\wh{y}(t+1)}+\E[t]{\wh{\pi}(t+1)} - \wh{i}(t).
\end{equation*} 
Last, we introduce the values of $\wh{y}(t)$ and $\wh{y}(t+1)$ given by \eqref{e:yhat}, and the value of $\wh{i}(t)$ given by~\eqref{e:ihat}. We obtain the IS equation:
\begin{equation}
\a \wh{n}(t)+\p \wh{\pi}(t)=\a \E[t]{\wh{n}(t+1)}+\E[t]{\wh{\pi}(t+1)} - \wh{i_0}(t) - \wh{a}(t)+ \E[t]{\wh{a}(t+1)}.
\label{e:is}\end{equation}

\paragraph{Short-Run Phillips Curve} The sixth and final equation is the short-run Phillips curve. It is based on the pricing equation~\eqref{e:ema}.

As a first step toward computing the Phillips curve, we compute the elasticity of the price elasticity of demand $E(M^p) = \e+(\e-1)\g\f(M^p)$. Given that the elasticity of $\f(M^p)$ is $\s$ (Lemma~\ref{l:phi}), the elasticity of $E(M^p)$ at the steady state is
\begin{equation}
\odl{E}{M^p} =  \frac{(\e-1)\g\ol{\f}}{\e+(\e-1) \g \ol{\f}}\cdot \ol{\s} \equiv \O_{0}.
\label{e:omega0}\end{equation}
Second, we introduce the auxiliary function 
\begin{equation*}
\L_1(M) = \frac{M-1}{M}. 
\end{equation*}
The elasticity of $\L_1(M)$ at the steady state is
\begin{equation*}
\odl{\L_1}{M} =\frac{\ol{M}}{\ol{M}-1}-1 = \frac{1}{\ol{M}-1}\equiv \O_{1}.
\end{equation*}
Using the value of $\ol{M}$ in \eqref{e:mssa}, we find that $\O_1$ satisfies 
\begin{equation}
\O_1 = (\e-1)\bs{1+\frac{(1-\d)\g}{1-\d\g}\ol{\f}}.
\label{e:omega1}\end{equation}
The left-hand side of \eqref{e:ema} can be written $LHS = \L_1(M(t)) \cdot E(M^p(t))$.
Accordingly, around the steady state the log-linear approximation of $LHS$ is
\begin{equation}
\ln(LHS)-\ln(\ol{LHS}) = \O_{1} \wh{m}(t) + \O_{0} \wh{m^p}(t).
\label{e:lhs}\end{equation}

Next, we introduce another auxiliary function: 
\begin{equation*}
\L_2(M^p) =E(M^p) - (1-\g)\e = \g \bs{\e +(\e-1)\f(M^p)}.
\end{equation*}
The elasticity of $\L_2(M^p)$ at the steady state is
\begin{equation}
\odl{\L_2}{M^p} =  \frac{(\e-1) \ol{\f}}{\e+(\e-1) \ol{\f}}\cdot \ol{\s} \equiv \O_{2}.
\label{e:omega2}\end{equation}
We also introduce the auxiliary function
\begin{equation*}
\L_3(x) = 1- \d\g + \d x,
\end{equation*}
whose elasticity is
\begin{equation*}
\odl{\L_3}{x} = \frac{\d x}{\L_3} \equiv \O_{3}.
\end{equation*}
The right-hand side of \eqref{e:ema} (abstracting from the expectation operator) can be written 
\begin{equation*}
RHS = \L_3(\L_1(M(t+1))\cdot \L_2(M^p(t+1)).
\end{equation*}
Hence, around the steady state the log-linear approximation of $RHS$ is
\begin{equation}
\ln(RHS)-\ln(\ol{RHS}) = \O_{3} \cdot  \bs{\O_{1} \wh{m}(t+1) + \O_{2} \wh{m^p}(t+1)},
\label{e:rhs}\end{equation}
where the elasticity $\O_3$ is evaluated at $\ol{x}=\ol{\L_1}\cdot \ol{\L_2}$ and $\ol{\L_3}= \ol{RHS} = \ol{LHS} =\ol{E} \cdot\ol{\L_1}$. Thus in \eqref{e:rhs} we have
\begin{equation}
\O_{3} = \frac{\d \ol{\L_1}\cdot \ol{\L_2}}{\ol{E}\cdot \ol{\L_1}} = \d\g \frac{\e +(\e-1)\ol{\f}}{\e+(\e-1) \g \ol{\f}}.
\label{e:omega3}\end{equation}

We now bring these results together. Equation \eqref{e:ema} can be written $LHS= \E[t]{RHS}$. This equation also holds in steady state so $\ol{LHS} =\ol{RHS}$. Combining these two equations, we infer
\begin{equation*}
\exp{\ln(LHS)-\ln(\ol{LHS})} = \E[t]{\exp{\ln(RHS)-\ln(\ol{RHS})}}.
\end{equation*}
Around $x=0$, we have $\exp(x) = 1 + x$. Applying this approximation to both sides of the previous equation, we find
\begin{equation*}
1+\ln(LHS)-\ln(\ol{LHS}) = 1+ \E[t]{\ln(RHS)-\ln(\ol{RHS})}.
\end{equation*}
We then use the results in \eqref{e:lhs} and \eqref{e:rhs}:
\begin{equation*}
\O_{1} \wh{m}(t) + \O_{0} \wh{m^p}(t) = \O_{3} \cdot  \bs{\O_{1} \E[t]{\wh{m}(t+1)} + \O_{2} \E[t]{\wh{m^p}(t+1)}}.
\end{equation*}
We divide this equation by $\O_{0}$; insert the values of $\wh{m}(t)$ and $\wh{m}(t+1)$ given by \eqref{e:mhat}; and insert the value of $\wh{m^p}(t+1)$ given by \eqref{e:mphata}. We obtain
\begin{equation}
-\frac{(1+\eta)\O_{1}}{\O_{0}} \wh{n}(t) + \wh{m^p}(t) = -\frac{(1+\eta)\O_{3}\O_{1}}{\O_{0}} \E[t]{\wh{n}(t+1)} + \frac{\g\O_{3}\O_{2}}{\O_{0}} \E[t]{\wh{\pi}(t+1)+\wh{m^p}(t)}.
\label{e:phillips0}\end{equation}
Using \eqref{e:omega0}, \eqref{e:omega1}, \eqref{e:omega2}, and \eqref{e:omega3}, we find that 
\begin{align*}
\frac{(1+\eta)\O_{1}}{\O_{0}} &= (1+\eta) \frac{\e+(\e-1) \g  \ol{\f}}{\g \ol{\f} \ol{\s}}\bs{1+\frac{(1-\d)\g}{1-\d\g}  \ol{\f}} \equiv	\l_1 \\
\frac{(1+\eta)\O_{3}\O_{1}}{\O_{0}}&= (1+\eta) \d \frac{\e+(\e-1)\ol{\f}}{\ol{\f} \ol{\s}} \bs{1+\frac{(1-\d)\g}{1-\d\g}  \ol{\f}} \equiv	\l_2\\
\frac{\g\O_{3}\O_{2}}{\O_{0}}&=\d\g^2 \frac{\e +(\e-1)\ol{\f}}{\e+(\e-1) \g \ol{\f}} \cdot \frac{(\e-1) \ol{\f}\ol{\s}}{\e+(\e-1) \ol{\f}} \cdot \frac{\e+(\e-1) \g \ol{\f}}{(\e-1)\g\ol{\f}\ol{\s}} = \d\g.
\end{align*}
Bringing these results into \eqref{e:phillips0}, we obtain the short-run Phillips curve:
\begin{equation}
(1-\d\g) \wh{m^p}(t) - \l_1  \wh{n}(t)  = \d \g  \E[t]{\wh{\pi}(t+1)} - \l_2  \E[t]{\wh{n}(t+1)}.
\label{e:phillipsa}\end{equation}

\paragraph{Proof of Lemma~\ref{l:monetary}} The law of motion \eqref{e:mphat} for the perceived price markup comes from \eqref{e:mphata}. The expression of the perceived price markup as a discounted sum of past inflation rates is obtained by iterating \eqref{e:mphata} backward; and by noting that $\lim_{T\to \infty}\g^T \cdot \wh{m^p}(t-T)=0$ as $\g\in(0,1)$ and $\wh{m^p}$ is bounded.

\paragraph{Proof of Proposition~\ref{p:monetary}} The short-run Phillips curve \eqref{e:phillips} comes from \eqref{e:phillipsa}. The hybrid expression of the short-run Phillips curve is obtained by combining \eqref{e:phillips} with \eqref{e:mphat}.

\paragraph{Blanchard-Kahn Representation} To complete the description of the log-linearized equilibrium, we combine the equilibrium conditions \eqref{e:mphata}, \eqref{e:is}, and \eqref{e:phillipsa} into a dynamical system of the form proposed by \ct{BK80}. Such system is useful to assess the existence and uniqueness of an equilibrium, and to solve for the unique equilibrium when it exists.

We first combine \eqref{e:mphata}, \eqref{e:is}, and \eqref{e:phillipsa} into a linear dynamical system:
\begin{equation*}
\bs{\begin{array}{ccc}
\g & \g & 0\\ 
0 & \p & \a \\ 
0  & 0 & \l_1 \\ 
\end{array}}
\bs{\begin{array}{c}
\wh{m^p}(t-1)\\
\wh{\pi}(t)\\
\wh{n}(t)\\
\end{array}}
=
\bs{\begin{array}{ccc}
1 &  0 & 0\\ 
0 & 1 & \a \\ 
1-\d\g & -\d\g & \l_2\\ 
\end{array}}
\bs{\begin{array}{c}
\wh{m^p}(t)\\
\E[t]{\wh{\pi}(t+1)}\\
\E[t]{\wh{n}(t+1)}\\
\end{array}}-
\bs{\begin{array}{c}
0\\
1\\
0\\
\end{array}} \o(t),
\end{equation*}
where 
\begin{equation*}
\o(t) = \wh{i_0}(t)+\wh{a}(t) - \E[t]{\wh{a}(t+1)}
\end{equation*}
is an exogenous shock realized at time~$t$. The inverse of the matrix on the right-hand side is
\begin{equation*}
\bs{\begin{array}{ccc}
1 &  0 & 0\\ 
0 & 1 & \a \\ 
1-\d\g  & -\d\g & \l_2\\ 
\end{array}}^{-1} = \bs{\begin{array}{ccc}
1 &  0 &  0\\
\frac{(1-\d\g)\a}{\l_2 + \a \d \g} &  \frac{\l_2}{\l_2 + \a \d \g} & \frac{-\a}{\l_2 + \a \d \g}\\
\frac{\d\g-1}{\l_2 + \a \d \g} & \frac{\d \g}{\l_2 + \a \d \g} &  \frac{1}{\l_2 + \a \d \g}\\
\end{array}}.
\end{equation*}
Premultiplying the dynamical system by the inverse matrix, we obtain the Blanchard-Kahn form of the system:
\begin{equation*}
\bs{\begin{array}{c}
\wh{m^p}(t)\\
\E[t]{\wh{\pi}(t+1)}\\
\E[t]{\wh{n}(t+1)}\\
\end{array}}
= \bs{\begin{array}{ccc}
\g & \g &  0\\
\frac{ (1 - \d \g )\a \g}{\l_2 + \a \d \g} & \frac{\l_2 \p + \a \g (1- \d \g)}{\l_2 + \a \d \g} & \frac{(\l_2-\l_1) \a}{\l_2 + \a \d \g}\\
\frac{-(1-\d \g)\g }{\l_2 + \a \d \g} & \frac{\bs{\d\p+\d\g- 1} \g}{\l_2 + \a \d \g} & \frac{\l_1 + \a \d \g}{\l_2 + \a \d \g}\\
\end{array}}
\bs{\begin{array}{c}
\wh{m^p}(t-1)\\
\wh{\pi}(t)\\
\wh{n}(t)\\
\end{array}}
+ \bs{\begin{array}{c}
0\\
\frac{\l_2}{\l_2 + \a \d \g}\\
\frac{\d \g}{\l_2 + \a \d \g}\\
\end{array}} \o(t).
\end{equation*}
This dynamical system determines perceived price markup $\wh{m^p}(t)$, inflation $\wh{\pi}(t)$, and employment $\wh{n}(t)$. All the other variables directly follow.

Under the calibration in Table~\ref{t:calibration}, the Blanchard-Kahn conditions are satisfied, so the equilibrium exists and is determinate. Indeed, under such calibration, the eigenvalues of the matrix in the Blanchard-Kahn system are $0.30$, $1.02+0.03i$, and $1.02-0.03i$: one eigenvalue is within the unit circle, and two are outside the unit circle. Further, the dynamical system has one predetermined variable at time~$t$ ($\wh{m^p}(t-1)$) and two nonpredetermined variables ($\wh{n}(t)$ and $\wh{\pi}(t)$). As the number of eigenvalues outside the unit circle matches the number of nonpredetermined variables, the solution to the dynamical system exists and is unique \cp[Proposition~1]{BK80}. 

\subsection{Calibration}\label{a:calibration}

Finally, we calibrate the fairness-related parameters of the New Keynesian model following the procedure described in Section~\ref{s:calibration}. The procedure is based on matching the cost passthroughs estimated in microdata and those obtained by simulating the behavior of a single firm facing a stochastic marginal cost. The calibrated values of the parameters are summarized in Table~\ref{t:calibration}.

\paragraph{Firm Problem} This is a simplified version of the New Keynesian firm problem, which abstracts from hiring decisions. The firm chooses price $P(t)$ and output $Y(t)$ to maximize the expected present-discounted value of profits
\begin{equation*}
\E[0] \sum_{t=0}^{\infty}\d^t \bs{P(t) - C(t)} Y(t),
\end{equation*}
 subject to the demand
\begin{equation}
Y^{d}(P(t),C^{p}(t-1)) = P(t)^{-\e}  F\of{\bp{\frac{\e}{\e-1}}^{1-\g}\bs{\frac{P(t)}{C^{p}(t-1)}}^{\g}}^{\e-1}
\label{e:ydjc}\end{equation}
and to the law of motion \eqref{e:cpj} for the perceived marginal cost $C^{p}(t)$. The nominal marginal cost $C(t)$ is exogenous and stochastic. 

To solve the firm's problem, we set up the Lagrangian:
\begin{align*}
\Lc& =\E[0]\sum_{t=0}^{\infty}\d^t \bigg\{ \bs{P(t) - C(t)} Y(t) \\
& + \Hc(t)  \bs{Y^{d}(P(t),C^{p}(t-1))-Y(t)}\\
&+\Kc(t)  \bs{\bs{C^{p}(t-1)}^{\g}  \bs{\frac{\e-1}{\e} P(t)}^{1-\g}-C^{p}(t)}\bigg\},
\end{align*}
where $\Hc(t)$ is the Lagrange multiplier on the demand constraint in period~$t$, and $\Kc(t)$ is the Lagrange multiplier on the perceived marginal cost's law of motion in period~$t$.

\paragraph{First-Order Condition with Respect to Output} The first-order condition with respect to $Y(t)$ is $\pdx{\Lc}{Y(t)}=0$. It yields
\begin{equation*}
\Hc(t)=P(t) \bs{1-\frac{C(t)}{P(t)}},
\end{equation*}
which is the same equation as \eqref{e:hj2} and thus can be rewritten as \eqref{e:hj}.

\paragraph{First-Order Condition with Respect to Price} The first-order condition with respect to $P(t)$ is $\pdx{\Lc}{P(t)}=0$, which gives
\begin{equation*}
0=Y(t)+\Hc(t)\pd{Y^{d}}{P}+(1-\g)\Kc(t)\frac{C^{p}(t)}{P(t)}.
\end{equation*}
This equation is the same as \eqref{e:focpj}; therefore, it can be re-expressed as \eqref{e:ejmj}.

\paragraph{First-Order Condition with Respect to Perceived Marginal Cost} Finally, the first-order condition with respect to $C^{p}(t)$ is $\pdx{\Lc}{C^p(t)}=0$, which yields
\begin{equation*}
0 =\d \E[t]{\Hc(t+1) \pd{Y^{d}}{C^{p}} +\g \Kc(t+1) \frac{C^{p}(t+1)}{C^{p}(t)}} -\Kc(t).
\end{equation*}
Using the elasticity given by \eqref{e:ydc}, we get
\begin{equation*}
\Kc(t) = \d \E[t]{\Hc(t+1) \frac{Y(t+1)}{C^{p}(t)} \bs{E(M^{p}(t+1))-\e} +\g \Kc(t+1) \frac{C^{p}(t+1)}{C^{p}(t)}}.
\end{equation*}
Next we multiply the equation by $C^p(t)/[Y(t)P(t)]$, and we insert the perceived price markups $M^p(t) = P(t)/C^{p}(t)$ and $M^p(t+1) = P(t+1)/C^{p}(t+1)$. We get
\begin{equation*}
\frac{\Kc(t)}{Y(t) M^{p}(t)}=\d \E[t]{\frac{Y(t+1) P(t+1)}{Y(t) P(t)}\bc{\frac{\Hc(t+1)}{P(t+1)} \bs{E(M^{p}(t+1))-\e} +\g\frac{\Kc(t+1)}{Y(t+1)M^{p}(t+1)}}}.
\end{equation*}
To conclude, we multiply the equation by $(1-\g)$, eliminate $\Hc(t+1)$ using \eqref{e:hj}, and eliminate $\Kc(t)$ and $\Kc(t+1)$ using \eqref{e:ejmj}. This gives the following pricing equation:
\begin{equation}\medmuskip=1mu\thickmuskip=2mu
\frac{M(t)-1}{M(t)} E(M^{p}(t))=1+\d \E[t]{\frac{Y(t+1) P(t+1)}{Y(t)P(t)}\bc{\frac{M(t+1)-1}{M(t+1)}\bs{E(M^{p}(t+1))-(1-\g)\e}-\g}}.
\label{e:emjc}\end{equation}
In steady state, this equation becomes \eqref{e:emss1} and can therefore be written as \eqref{e:mssa}.

\paragraph{Firm Pricing} The firm's pricing is described by four variables: the price $P(t)$, markup $M(t)$, output $Y(t)$, and perceived markup $M^p(t)$. These four variables are determined by four equations: the pricing equation~\eqref{e:emjc}, the definition of the markup $M(t) = P(t)/C(t)$, the demand curve~\eqref{e:ydjc}, and the perceived markup's law of motion~\eqref{e:mpt}.

\begin{figure}[t]
\includegraphics[scale=0.3,page=15]{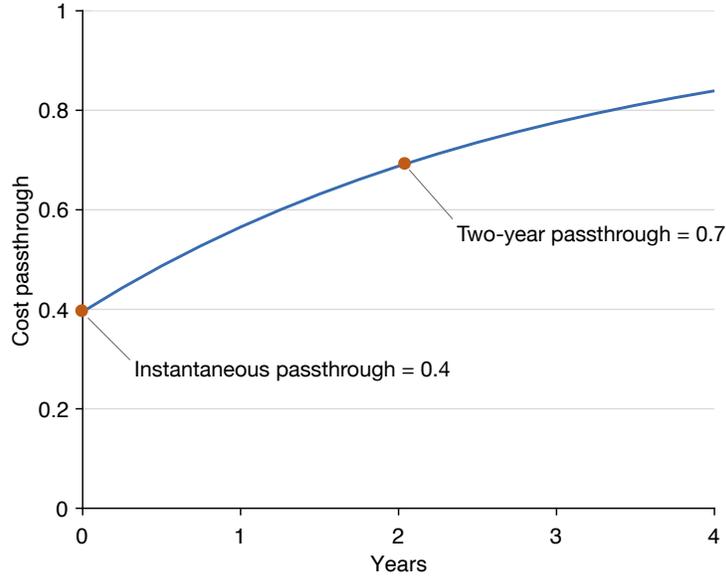}
\caption{Simulated dynamics of cost passthrough}
\note{The cost passthrough represents the percentage increase in price due to a 1\% increase in marginal cost. The empirical estimates of the cost passthrough ($0.4$ and $0.7$) are obtained in Section~\ref{s:calibration}. The simulations are obtained from the pricing model in Appendix~\ref{a:calibration} under the calibration in Table~\ref{t:calibration}.}
\label{f:calibration}\end{figure}

\paragraph{Simulations} We start from a steady-state situation. To be consistent with the simulations of Figures~\ref{f:monetary} and~\ref{f:technology}, we assume that steady-state inflation is zero, so the marginal cost $C$ is constant in steady state. Then we impose an unexpected permanent 1\% increase in $C$. We compute the firm's response to this shock by solving the nonlinear dynamical system of four equations that describes firm's pricing. We obtain the dynamics of the cost passthrough by calculating the percentage change in price over time: 
\begin{equation*}
\b(t) = \frac{P(t)-\ol{P}}{\ol{P}} \times 100.
\end{equation*}

\paragraph{Calibration Procedure} We set the shape of the fairness function to \eqref{e:f} and the discount factor to $\d=0.99$. Then, using the simulations, we calibrate the three main parameters of the model: the concern for fairness, $\t$, the degree of underinference, $\g$, and the elasticity of substitution between goods, $\e$. Our goal is to produce an instantaneous cost passthrough of $\b = 0.4$, a two-year cost passthrough of $\b =0.7$, together with a steady-state price markup of $\ol{M}=1.5$.

Our calibration procedure starts by initializing $\t$ and $\g$ to some values. Using these values and the target $\ol{M}=1.5$, we compute $\e$ from \eqref{e:mssa}. In \eqref{e:mssa} we use \eqref{e:phia}, which holds because the fairness function is \eqref{e:f}, and because there is no inflation in steady state so customers are acclimated. Using the values of $\t$, $\g$, and $\e$, we simulate the dynamics of the cost passthrough. 

We repeat the simulation for different values of $\t$ and $\g$ until we obtain a cost passthrough of $0.4$ on impact and $0.7$ after two years. We reach these targets with $\t=9$ and $\g=0.8$. The corresponding value of $\e$ is $2.23$. The corresponding passthrough dynamics are shown in Figure~\ref{f:calibration}.

\section{Textbook New Keynesian Model}\label{a:textbook}

We describe the textbook New Keynesian model used as benchmark in the simulations of Figures~\ref{f:monetary} and~\ref{f:technology}. The model originates in \ct{G08}. The pricing friction in the model is the staggered pricing of \ct{C83}. Some researchers alternatively use the price-adjustment cost of \ct{R82}. However, since both pricing frictions yield the same linearized Phillips curve around the zero-inflation steady state, simulations in the two cases coincide \cp[pp.~976--979]{R95}.

The model's dynamics around the zero-inflation steady state are governed by an IS equation and a short-run Phillips curve. The IS equation is given by \eqref{e:is}, as in the model with fairness. This IS equation is obtained from (12) in \ct[Chapter~3]{G08}, by using logarithmic consumption utility, and by incorporating the monetary-policy rule~\eqref{e:ihat} and the production function~\eqref{e:yhat}.

The short-run Phillips curve is given by
\begin{equation*}
\wh{\pi}(t)=\d \E[t]{\wh{\pi}(t+1)} + \k \wh{n}(t),
\end{equation*}
where
\begin{equation*}
\k \equiv (1+\eta) \cdot \frac{(1-\x)(1-\d\x)}{\x} \cdot \frac{\a}{\a+\bp{1-\a} \e},
\end{equation*}
and $\x$ is the fraction of firms keeping their prices unchanged each period. This Phillips curve is obtained from (21) in \ct[Chapter~3]{G08}, by using logarithmic consumption utility, and by replacing the output gap by $\a \wh{n}(t)$.\footnote{The output gap is the logarithmic difference between the actual and the natural level of output. The natural levels of output and employment are reached when prices are flexible, so when the price markup is $\e/(\e-1)$. Since $\e/(\e-1)$ is also the steady-state price markup, we infer from \eqref{e:na} that the natural and steady-state levels of employment are equal. Hence, \eqref{e:y} implies that the natural level of output is $Y^n(t)= A(t) \ol{N}^{\a}$. Consequently the output gap is $\ln(Y(t))- \ln(Y^n(t))= \a [\ln(N(t))-\ln(\ol{N})] = \a\wh{n}(t)$.}

The IS equation and short-run Phillips curve jointly determine employment $\wh{n}(t)$ and inflation $\wh{\pi}(t)$. The other variables directly follow from $\wh{n}(t)$ and $\wh{\pi}(t)$. The nominal interest rate $\wh{i}(t)$ is given by~\eqref{e:ihat}. Output $\wh{y}(t)$ is given by~\eqref{e:yhat}. The price markup $\wh{m}(t)$ is given by~\eqref{e:mhat}. Since households observe both prices and costs, perceived and actual price markups are equal: $\wh{m^p}(t) = \wh{m}(t)$.

\end{document}